\documentclass[aps,prd,12pt,notitlepage]{revtex4}
    \usepackage{graphicx}
    \usepackage{amsmath,amssymb,mathrsfs}
\usepackage{epsf}
\usepackage{slashed}
\usepackage{epsfig}
\usepackage[font=small,skip=0pt,justification=raggedright]{caption}

\newcounter{fig}

\newcommand{\bea}{\begin{eqnarray}}
\newcommand{\eea}{\end{eqnarray}}
\newcommand{\be}{\begin{equation}}
\newcommand{\ee}{\end{equation}}

\newcommand{\re}[1]{(\ref{#1})}

\begin{document}

\title{Q-balls without a potential}

\author{V.~Loiko$^{\star}$, I. Perapechka$^{\star}$ and Ya.~Shnir$^{\dagger}$
}
\affiliation{$^{\star}$Department of Theoretical Physics and Astrophysics,
Belarusian State University, Minsk 220004, Belarus\\
$^{\dagger}$BLTP, JINR, Dubna 141980, Moscow Region, Russia
}

\begin{abstract}
We study non-topological Q-ball solutions of the (3+1)-dimensional Friedberg-Lee-Sirlin two-component
model. The limiting case of vanishing potential term yields an example of  hairy Q-balls, which
possess a long range massless real field.
We discuss the properties of these stationary field configurations and determine their domain of existence.
Considering Friedberg-Lee-Sirlin model we present numerical evidence for the existence of
spinning axially symmetric Q-balls with different parity.
Solution of this type exist also in the limiting case of vanishing scalar potential.
We find that the hairy Q-balls are classically stable for all range of values of angular frequency.

\end{abstract}
\maketitle

\section{Introduction}
Q-balls are stationary non-topological solitons, which may exist in a model with complex scalar fields.
Configurations of this type were introduced in 1976 by
Friedberg, Lee and Sirlin in two-component model with symmetry breaking potential \cite{Friedberg:1976me}, a few years later
Coleman found another realization of Q-balls
considering a single complex scalar field
in a model with a non-renormalizable self-interaction potential \cite{Coleman:1985ki}.
In both cases the global phase invariance of the scalar field is associated with a conserved
$U(1)$ Noether charge $Q$, the Q-balls correspond to stationary points of the total energy functional  for a given value of the charge.

Q-balls attracted a lot of attention because it was suggested that such configuration may contribute to various scenario of
the evolution of the early Universe \cite{Friedberg:1986tq,Jetzer:1991jr,Lee:1986ts}.
The Q-balls also can occur in the minimal supersymmetric generalization of the Standard Model, where
one finds leptonic and baryonic Q-balls related with conservation of  lepton and baryon number, respectively
\cite{Kuzenko}. Further, it was argued that these Q-balls may play an essential role in baryogenesis via the Affleck-Dine
mechanism \cite{Affleck}, they also were considered as candidates for dark matter \cite{Kusenko:1997si}.
Q-balls may exist in a wide variety of physical systems. Solutions of
that type were constructed in Abelian gauge models with local $U(1)$ symmetry \cite{Anagnostopoulos:2001dh,Lee:1988ag},
in non-Abelian gauge theories \cite{Friedberg:1976me,Kusenko:1997vi,Safian:1987pr} and other models.
An interesting
realization of the Q-balls exists in  condensed matter systems where they appear in the Bose-Einstein condensate
\cite{Enqvist}, or in the superfluid $^3$He-B \cite{Bunkov}.

Spherically symmetric Q-balls exist only within a certain angular frequency range, which is determined by
the explicit structure of the potential.
Notably, in the Friedberg-Lee-Sirlin model, which describes dynamics of a real self-interacting scalar field,
coupled to a complex scalar field, the lower critical frequency is zero.
Typically, there are two branches of Q-ball solutions, which are represented by two curves of the dependencies of
the energy of the configuration on its charge \cite{Friedberg:1976me}. Solutions are stable along the
lower branch, when their mass is smaller than the mass of free charged quanta of scalar excitations.
In the simplest case the Q-balls are spherically symmetric, however there are generalized spinning axially symmetric
solutions with non-zero angular momentum \cite{Volkov:2002aj,Kleihaus:2005me}. The energy density of these spinning Q-balls
is of toroidal shape.

There are some important differences between the Q-ball solutions of the  Coleman model \cite{Coleman:1985ki} with a
single complex field and sextic potential, and the corresponding solutions of
the renormalizable Friedberg-Lee-Sirlin model \cite{Friedberg:1976me}.
An interesting feature of Q-balls in the
Friedberg-Lee-Sirlin model is that in 3+1 dimensions these localized configurations with finite energy
may also exist in the limiting case of vanishing scalar potential \cite{Levin:2010gp}.
It was pointed out that in such a limit the Q-balls are stabilized by the gradient terms in the energy functional.
Further, the corresponding real scalar component becomes massless, it possess Coulomb-like asymptotic tail.

However, the paper \cite{Levin:2010gp} contains only qualitative discussion of the corresponding solutions, they were
not constructed explicitly, further the authors
do not study the frequency dependence of these solitons and their stability.

The main purpose of this work is to construct explicit examples of stationary solutions of the Friedberg-Lee-Sirlin model in the
limit of vanishing potential that have not been studied so far,
fully investigate  properties of these Q-balls and determine their domains of existence.
We also consider spinning configurations with non-zero angular momentum with both even and odd parity
and address the issue of classical stability of these solutions.

\section{Spherically symmetric solutions}

The 3+1 dimensional Friedberg-Lee-Sirlin model describes a real self-interacting scalar field $\xi$, coupled to a
complex scalar field $\phi$:
\be
L= (\partial_\mu\xi)^2 + |\partial_\mu\phi|^2 - m^2 \xi^2|\phi|^2 - U(\xi) \, ,
\label{lag-fls}
\ee
where $\kappa$ is the coupling constant.  The potential of the real scalar field is
\be
U(\xi)= \mu^2 (1-\xi^2)^2 \, ,
\ee
thus, $\xi \to 1$ in the vacuum and the complex field $\phi$ becomes massive due to the coupling
with its real partner. Thus, the parameters  $\mu$ and $m$ corresponds to the
mass of the real and complex components, respectively.

Similar to the Coleman model with a non-renormalizable sextic potential \cite{Coleman:1985ki}, the model \re{lag-fls} is
invariant under the global $U(1)$ transformations of the complex field $\phi \to \phi e^{i\alpha}$.
The Noether current, associated with this symmetry, is
\be
j_{\mu} = i \left(\phi \partial_\mu \phi^*  - \phi^* \partial_\mu \phi \right); \qquad \partial^\mu j_{\mu}=0 \, ,
\label{Nother-curr}
\ee
and the conserved charge is
\be
Q=i\int d^3x \left(\phi \partial_t\phi^* - \phi^* \partial_t \phi\right) \, .
\label{Nother-charge}
\ee

First, we consider the usual spherically symmetric parametrization of the fields
\be
\xi=X(r)\, ;\qquad \phi= Y(r)e^{i\omega t} \, ,
\label{ansatz}
\ee
where $X(r)$ and $Y(r)$ are real functions of radial variable and $\omega$ is the frequency of stationary rotation.
\begin{figure}[h!]
\begin{center}
\setlength{\unitlength}{0.1cm}
\includegraphics[height=.26\textheight, angle =0]{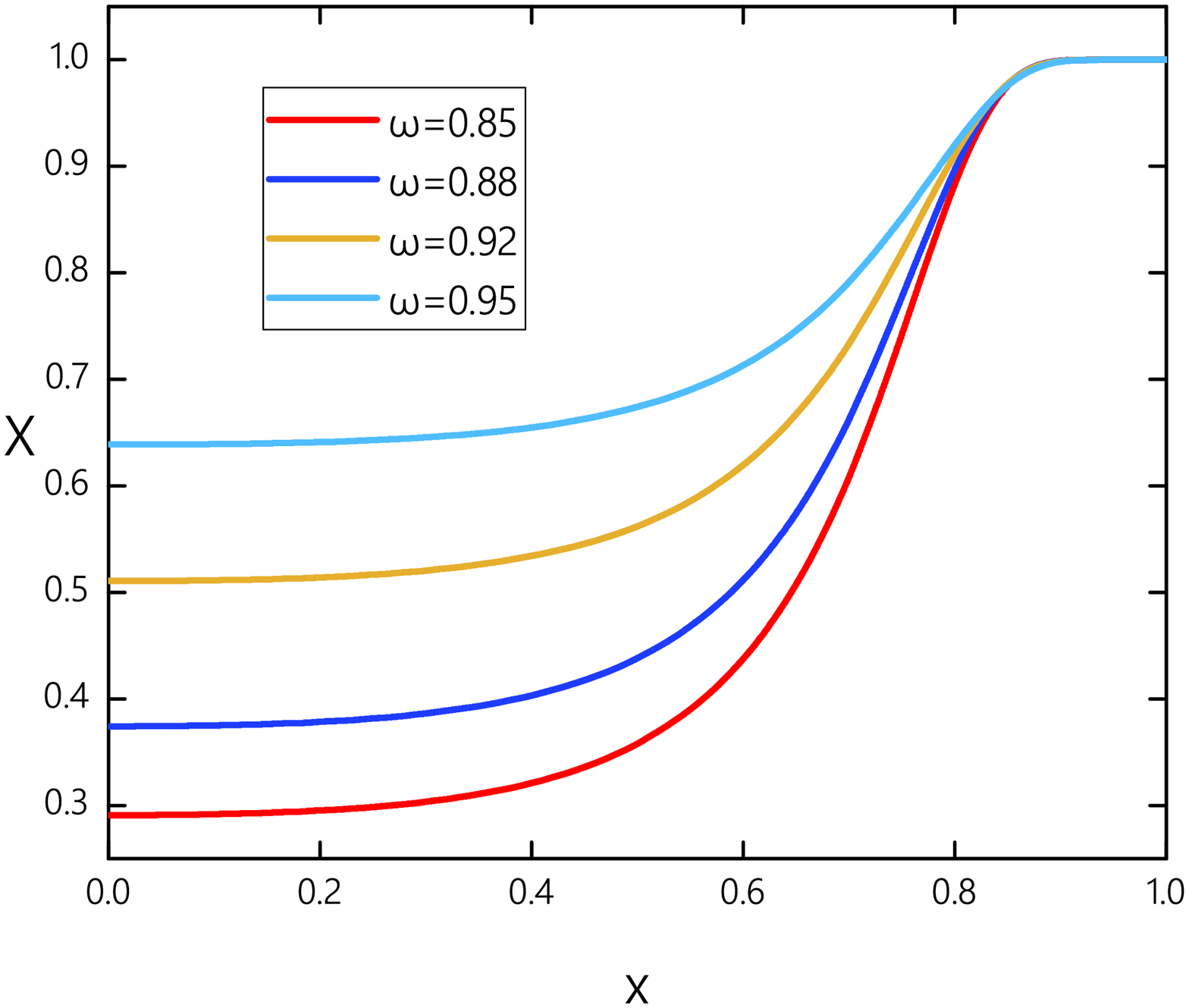}
\includegraphics[height=.26\textheight, angle =0]{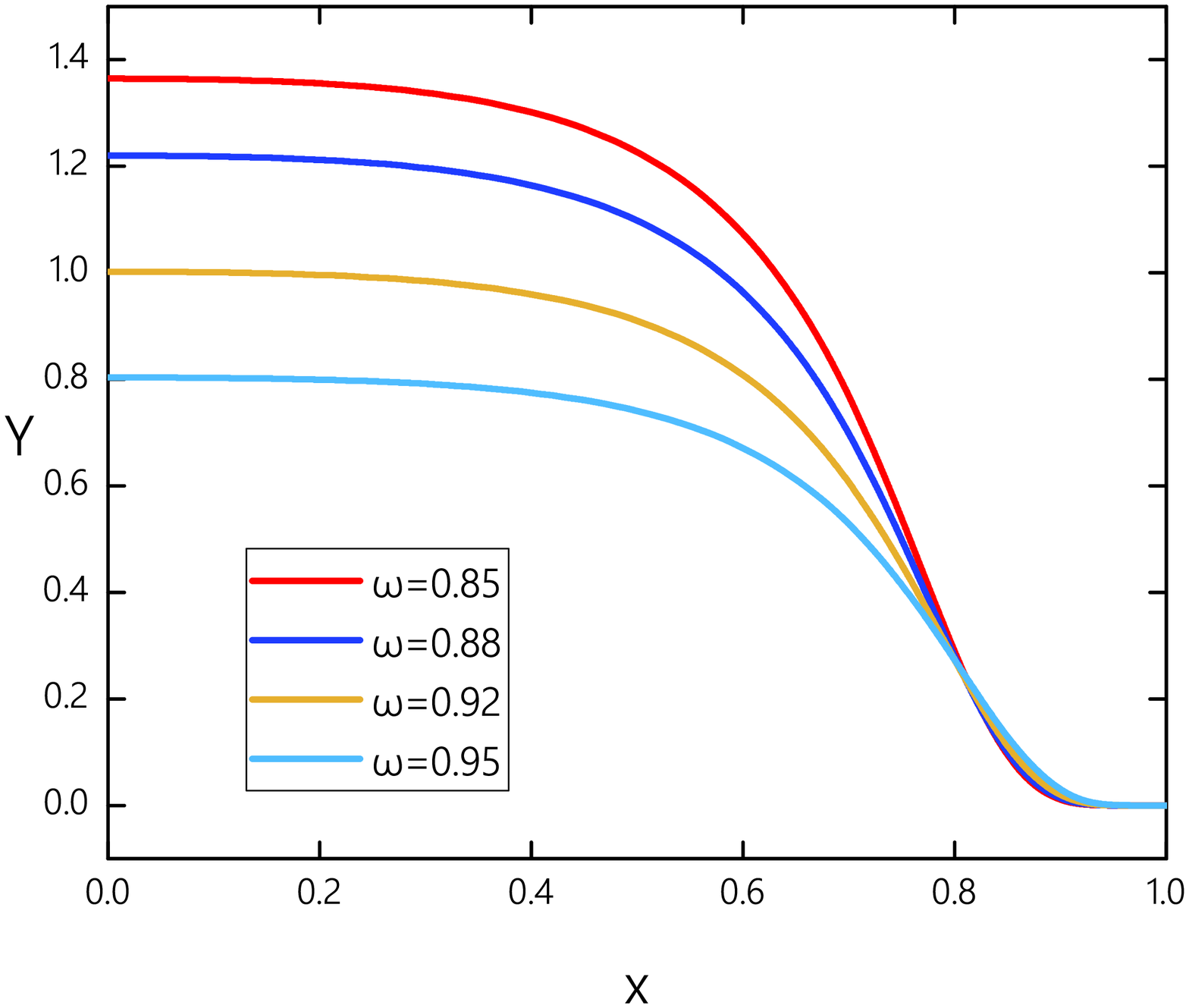}
\includegraphics[height=.26\textheight, angle =0]{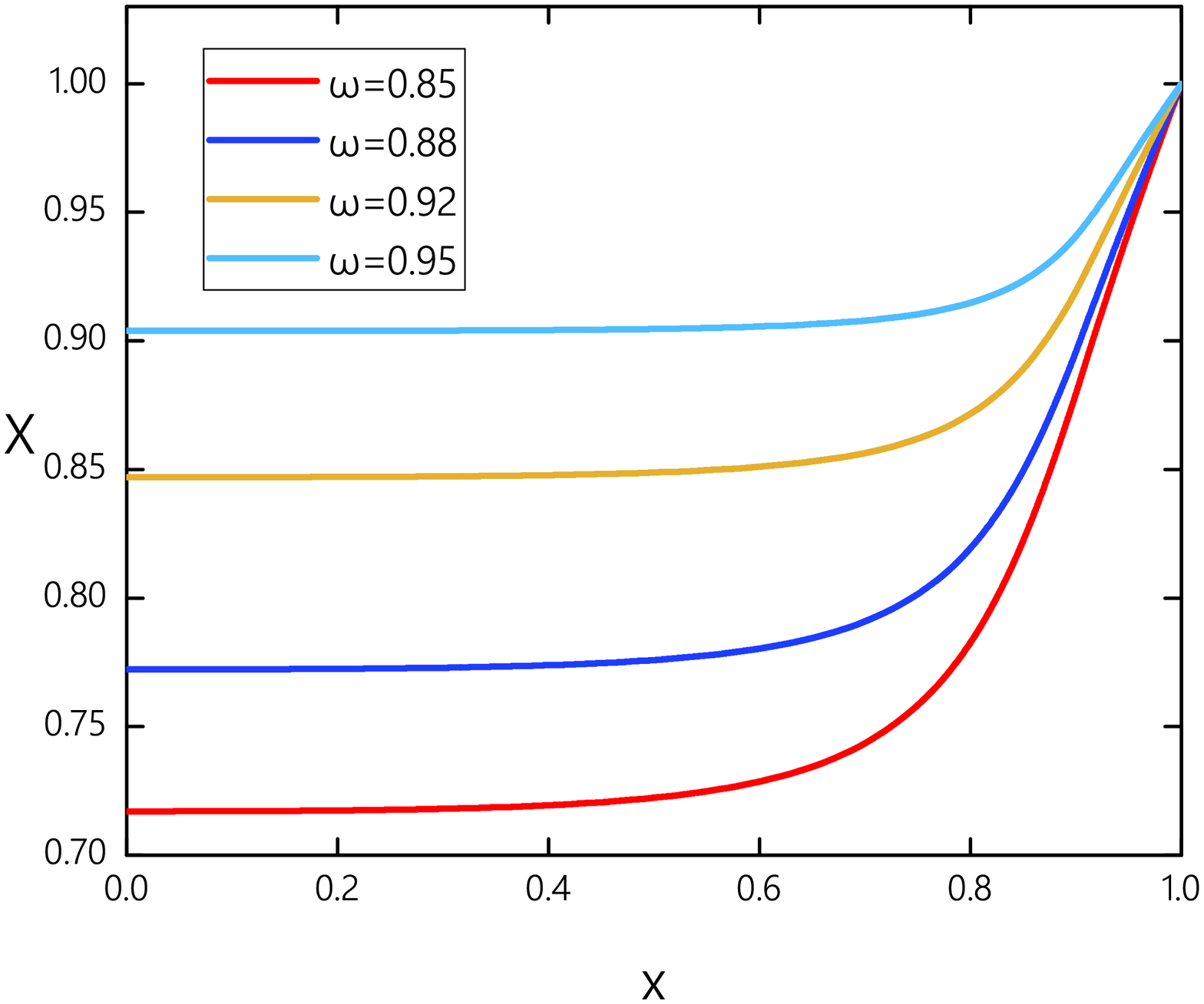}
\includegraphics[height=.26\textheight, angle =0]{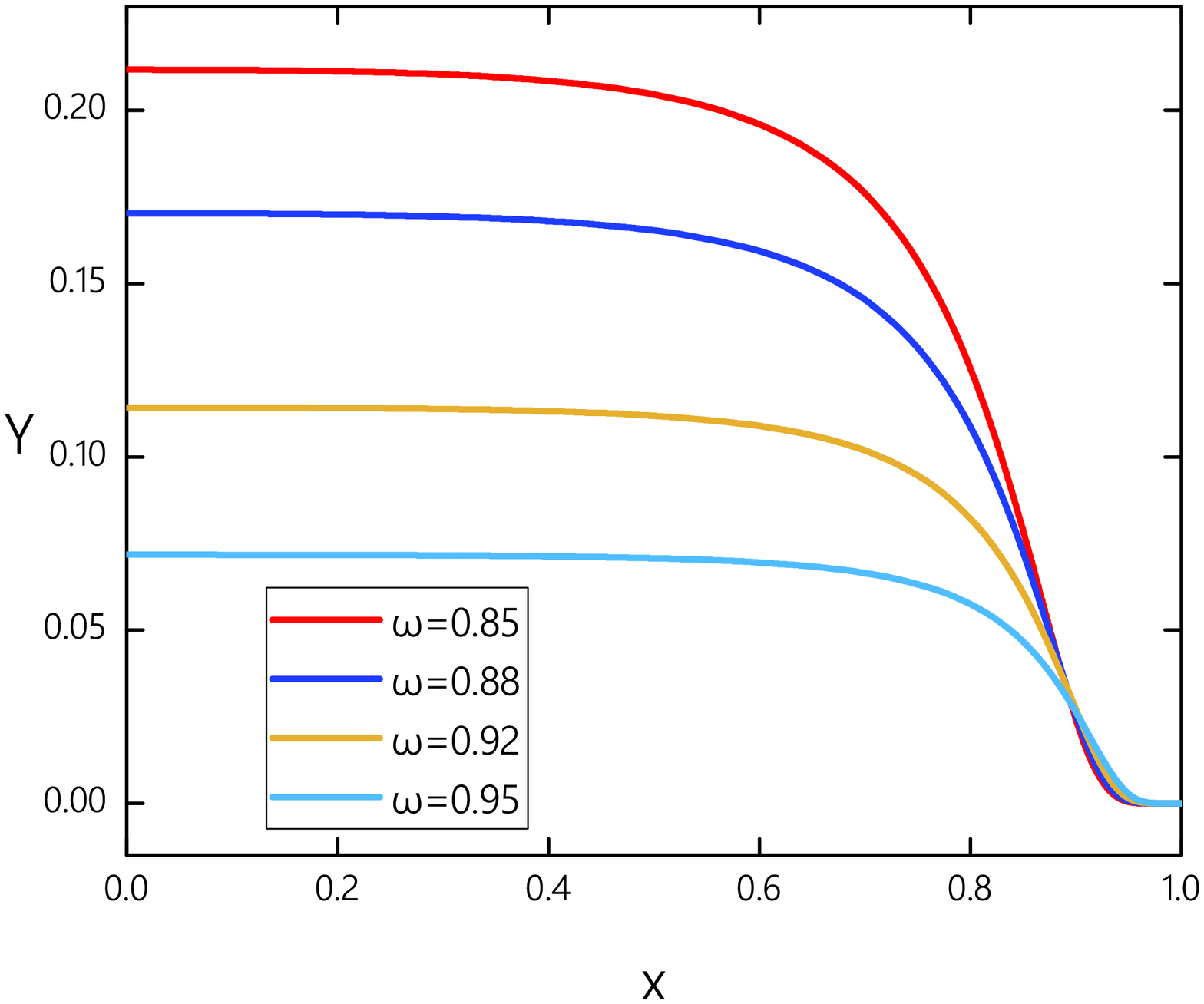}
\end{center}
\caption{The profiles of the field components of the Friedberg--Lee--Sirlin Q-ball configuration are plotted as
functions of the compact radial variable $x=r/(1+r)$ for some set of values
of angular frequency $\omega$ at $m=1$ and $\mu^2=0.25$ (upper plots) and $\mu^2=0$ (bottom plots).}
\label{fig1}
\end{figure}

Substitution of this ansatz into the stationary energy functional gives
\be
E= \int d^3 x \, T_{0}^0 = 4\pi \int\limits_{0}^\infty\!\! dr r^2 \left[\left( \frac{d X}{dr}\right)^2 +
\left( \frac{d Y}{dr}\right)^2 + \omega^2 Y^2 + \mu^2(1-X^2)^2 + m^2 Y^2 X^2 \right] \, ,
\label{Q-eng}
\ee
and  the charge of the spherically symmetric Q-ball is
\be
\label{Q-charge}
Q=8\pi\omega \int\limits_{0}^\infty\!dr ~r^2 Y^2 \, .
\ee

The field equations of the model become
\be
\begin{split}
\frac{d^2 X}{dr^2} &+  \frac{2}{r} \frac{dX}{dr} + 2 \mu^2 X (1-X^2) - m^2 X Y^2   =0\, ;\\
\frac{d^2 Y}{dr^2} &+  \frac{2}{r} \frac{dY}{dr} + \omega^2 Y - m^2 X^2 Y   =0\, .
\end{split}
\label{eqs}
\ee


Like other Q-ball dynamical equations \cite{Friedberg:1976me,Volkov:2002aj,Radu:2008pp,Brihaye:2007tn,shnir},
this system effectively describes a a unit mass pseudo-particle moving in two dimensional plane parameterized by
the "coordinates" $X(r),Y(r)$ and "time" $r$, in direction $Y$ in the effective potential
\be
U_{eff} = \omega^2 Y^2 - \mu^2(1-X^2)^2 - m^2 Y^2 X^2
 \,
 \label{eff-pot}
\ee
Non-topological soliton solution may exist when the trivial
configuration $X=Y=0$ corresponds
to a local maximum of the effective potential. This restriction corresponds to the upper bound on the angular frequency
$\omega_+ = m$, the Q-ball continuously approaches perturbative solutions as $\omega$ approaches this critical value
\cite{Friedberg:1976me}. Thus, the upper bound of the angular frequency of the Q-ball corresponds to the
mass of the complex component $m$, the localized soliton configuration with finite energy may exist as its
real component $X(r)$ becomes massless \cite{Levin:2010gp}. Hereafter we fix $m=1$, without loss of generality.
Note that, unlike the Q-balls in the non-renormalizable model with single complex field and sextic potential \cite{Coleman:1985ki},
there is no lower bound on the frequency, the solutions of the model \re{lag-fls} exists for all non-zero values of the angular
frequency $\omega$. As $\omega$ decreases, the characteristic size of the configuration is increasing.

The vacuum boundary conditions on the spacial infinity are $X\to 1$, $Y\to 0$ as $r\to \infty$ and the condition
of regularity
at the origin is
$$
\frac{d X}{dr} = \frac{d Y}{dr} =   0\, ,\quad  {\rm as} ~~~ r\to 0 \, .
$$
Imposing the boundary conditions we can find numerical solutions of
the system of coupled ordinary differential equations \re{eqs}. In our calculations we used
the usual shooting algorithm, based on Dormand-Prince 8th order method with
adaptive stepsize. The relative errors of calculations are lower than $10^{-10}$.

In Fig.~\ref{fig1} we displayed the corresponding solutions
at $m=1$ and $\mu^2=0.25$.
The parameter $\mu$ yields the mass of the excitations of the real component  $X$, it approaches the vacuum asymptotic value
as $X \sim 1- e^{- \sqrt{\mu^2-\omega^2} ~ r } $. Setting $\mu=0$ changes the asymptotic behavior, in such a case the real massless
field $X(r)$ decays
as   \cite{Levin:2010gp}
\be
X(r) \sim 1 - \frac{C}{r} + ~O(r^{-2}) \qquad {\rm ~~ as} ~~ r\to \infty
\label{massless-asymp}
\ee
This is a long-range Coulomb asymptotic with a scalar charge $C$, see Fig.~\ref{fig1}, bottom plots.

\begin{figure}[h!]
\begin{center}
\setlength{\unitlength}{0.1cm}
\includegraphics[height=.26\textheight, angle =0]{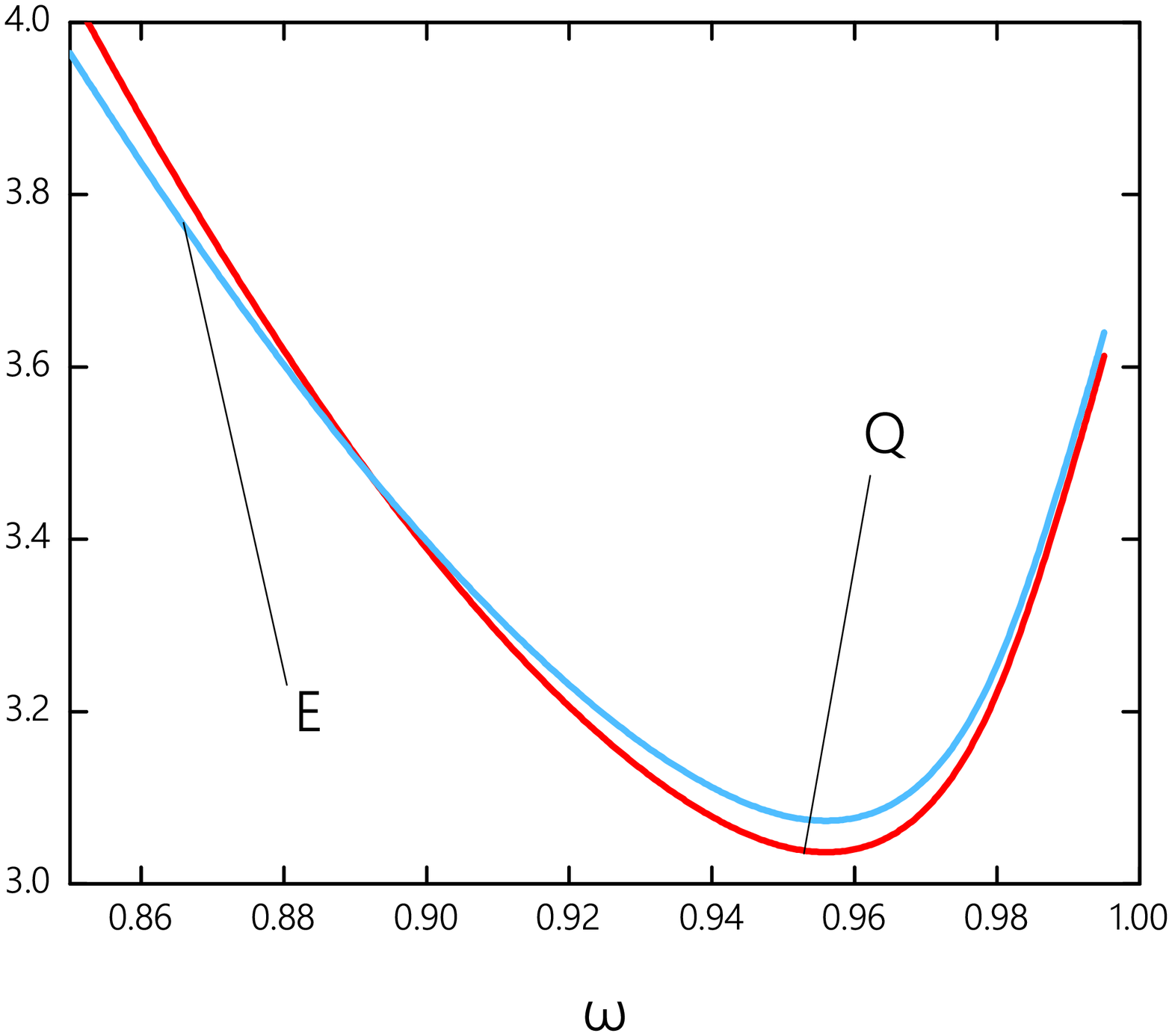}
\includegraphics[height=.26\textheight, angle =0]{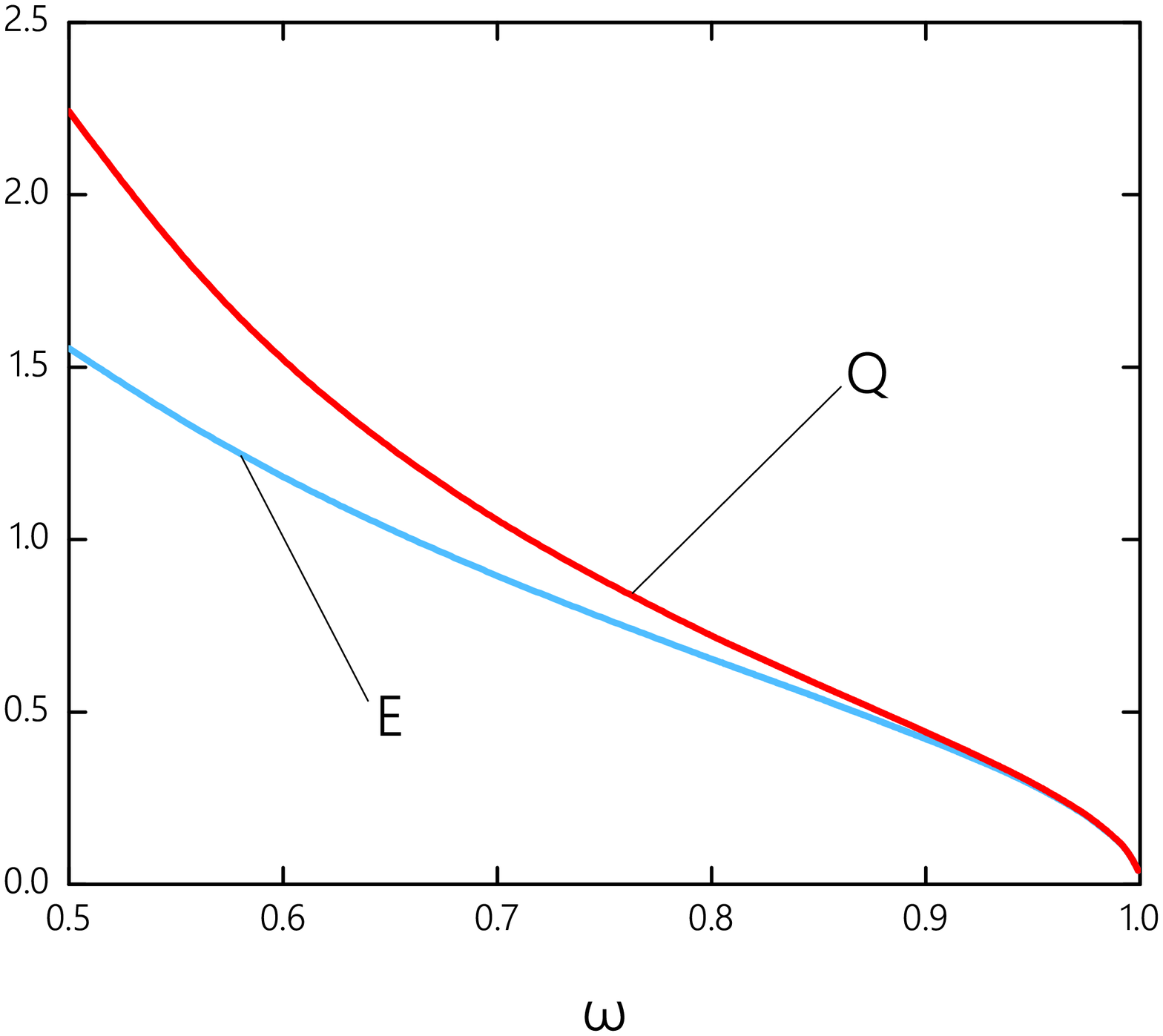}
\end{center}
\caption{The energy and the charge of the spherically symmetric Q-balls are shown in units of $16\pi$  as functions
of the angular frequency $\omega$ at $m=1$, $\mu^2=0.25$ (left) and $\mu^2=0$ (right).
}
\label{fig2}
\end{figure}

\begin{figure}[h!]
\begin{center}
\setlength{\unitlength}{0.1cm}
\includegraphics[height=.26\textheight, angle =0]{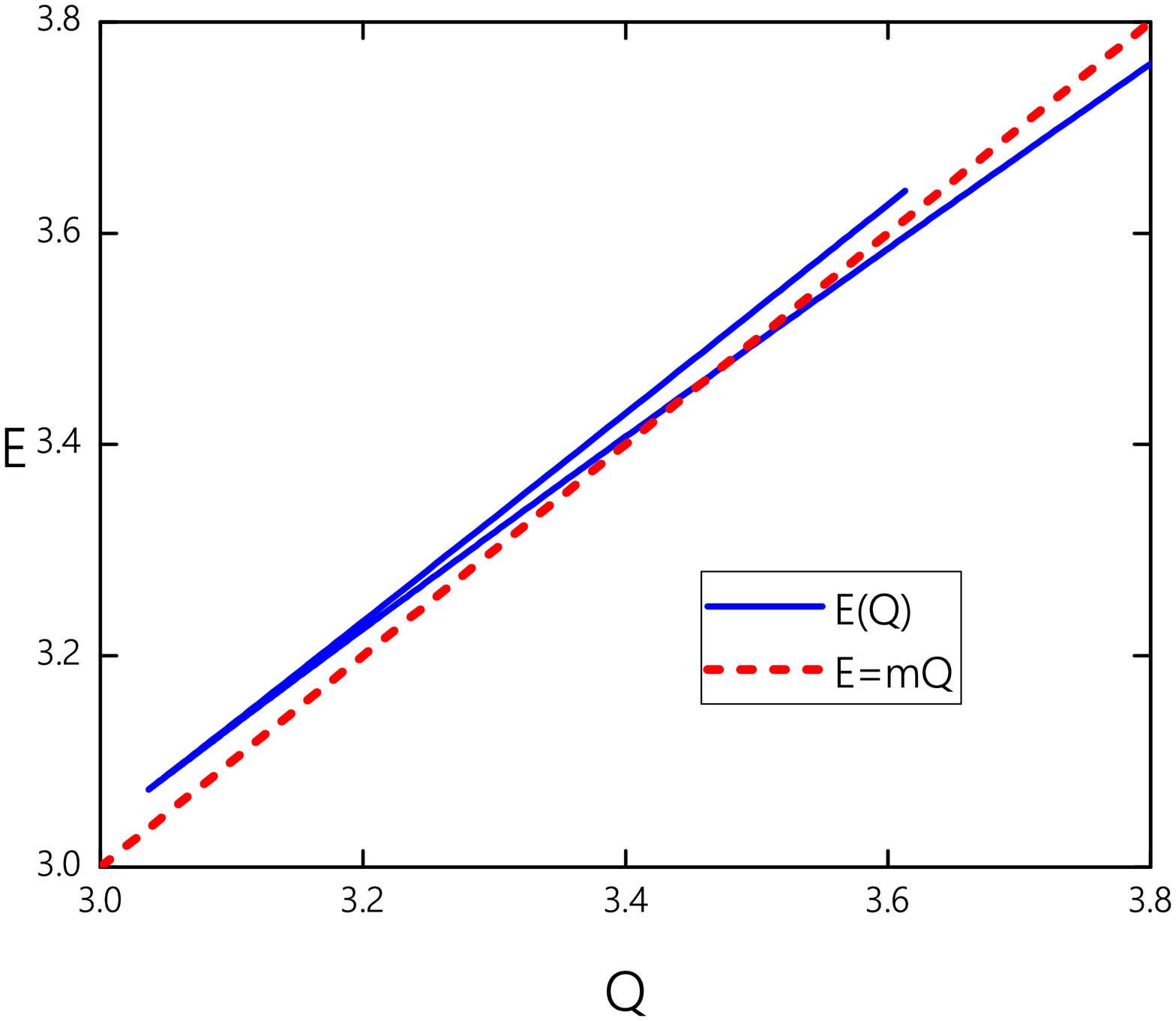}
\includegraphics[height=.26\textheight, angle =0]{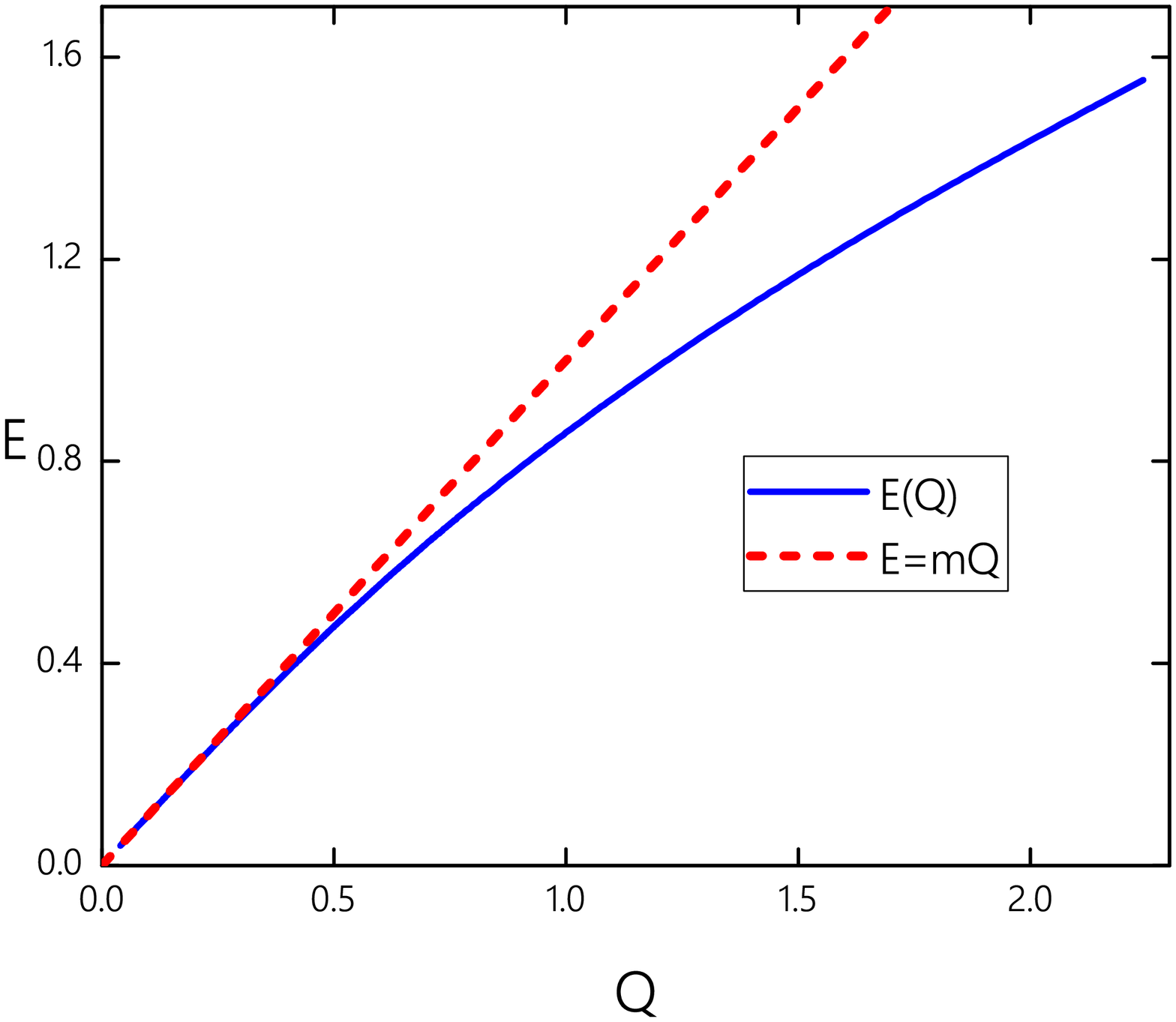}
\end{center}
\caption{The energy of the spherically symmetric Q-balls is plotted as a function of the charge $Q$
at $m=1$, $\mu^2=0.25$ (left) and $\mu^2=0$ (right). The straight line $E=mQ$ indicates the boundary of
the stability region.
}
\label{fig3}
\end{figure}

The charge and the energy of the Q-balls are given by \re{Q-eng},\re{Q-charge}, respectively. As the mass of the real component
remains non-zero, the curves of dependency  of both the charge and the mass  on the angular frequency exhibit the
typical two-branch behavior, see Fig.~\ref{fig2}, left plot. At some critical value of the frequency $\omega_{cr}$
the energy and the charge of the configuration are taking minimal non-zero values, they
diverge as $\omega$ approaches the upper bound, and in the opposite limit, as $\omega$ decreases to zero. Decrease of the mass parameter
$\mu$ for a fixed value of the frequency $\omega$ yields decrease of both the energy and the charge of the Q-ball.
Fig.~\ref{fig3} exhibits the $E(Q)$ curves of
the configurations at $\mu^2=0.25$ and in the massless case $\mu=0$.
In Fig.~\ref{fig3} we also indicate the energy of $Q$ free scalar quanta $E=mQ$, this
is a straight line separating the stability region, the configuration is classically stable below this line.

Indeed, for $\mu \neq 0$, there are
two branches of $E(Q)$ curves with a sharp cusp at $\omega= \omega_{cr}$
(see Fig.~\ref{fig3}, left plot).
The lower in energy branch corresponds to the values of the frequency $\omega < \omega_{cr}$ whereas the
upper branch corresponds to $\omega > \omega_{cr}$.
As the real component remains massive, the
configurations on the upper branch are unstable \cite{Friedberg:1976me}. We observe that decrease of
the mass parameter $\mu$ shifts the critical
value $\omega_{cr}$ towards the upper limit $\omega_+$.

The situation changes dramatically in the massless limit $\mu =0 $, the stable branch extends all the way up to the
critical value $\omega_+ = m$, here both components of the Q-ball approach the corresponding vacuum values and both the energy
and the charge of the configuration tend to zero, see Fig.~\ref{fig2}. Thus, the hairy Q-balls with long-range real scalar component
are classically stable for all range of values of the angular frequency.

\begin{figure}[h!]
\begin{center}
\setlength{\unitlength}{0.1cm}
\includegraphics[height=.34\textheight, angle =0]{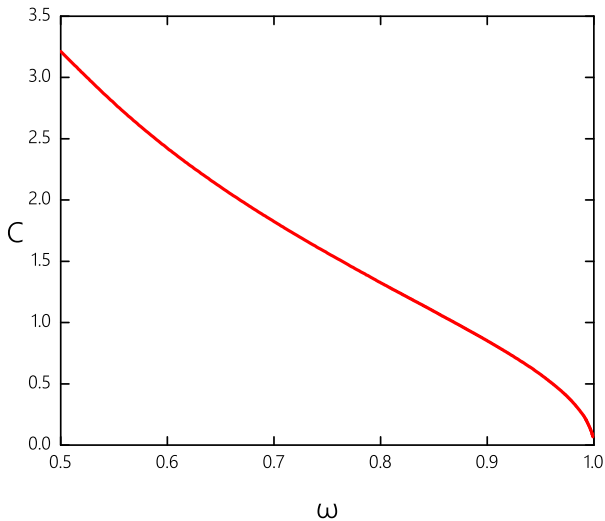}
\end{center}
\caption{The scalar charge of the  Q-balls with massless component $X$ is plotted as a function of the
frequency $\omega$.}
\label{fig8}
\end{figure}

Note that the scalar charge $C$, which corresponds to the Coulomb asymptotic tail of the hairy Q-ball \re{massless-asymp}, is not a constant.
The configuration is not rigid, its characteristic size varies with the angular frequency $\omega$. Indeed, numerical calculations show that
the value of the charge $C$ monotonically decreases, as $\omega$ increases, see Fig.~\ref{fig8}. However, the relation $C=\sqrt{Q/2}$,
which follows from some qualitative arguments \cite{Levin:2010gp},
holds only for large Q-balls, i.e. for small values of the angular frequency $\omega$.


\section{Spinning solutions}

A generalization of the fundamental spherically symmetric Q-ball can be constructed when we
consider spinning axially symmetric configuration \cite{Volkov:2002aj,Kleihaus:2005me}:
\be
\label{ansatz-spin}
\xi=X(r,\theta)\, ;\qquad \phi= Y(r,\theta)e^{i(\omega t + n\varphi)} \, ,
\ee
which generalizes the ansatz \re{ansatz}. Here $n \in \mathbb{Z}$ is a rotational quantum number, the angular
momentum of the stationary spinning Q-ball is classically quantized as \cite{Volkov:2002aj}
\be
J = \int d^3 x \, T_{\varphi}^0 = 4\pi n \omega \int\limits_0^\infty \int\limits_0^\pi dr d\theta \! ~\sin\theta ~ r^2 Y^2
 = n Q \, .
\ee
The real functions
$X(r,\theta),Y(r,\theta)$ depend on the polar angle $\theta$ and radial variable $r$.
The energy of the configuration then read
\be
\begin{split}
E &= 2 \pi \int\limits_0^\pi\int\limits_0^\infty\!  \sin\theta r^2 dr d\theta
\biggl[ X_r^2 + Y_r^2 + \frac{X_\theta^2}{r^2} + \frac{Y_\theta^2}{r^2} +
\left(\omega^2
+ \frac{n^2}{r^2 \sin^2 \theta}\right)Y^2\\
&+  \mu^2(1-X^2)^2 + m^2 Y^2 X^2 \biggr] \, .
\end{split}
\label{Q-eng-2}
\ee
The corresponding field equations are
\be
\begin{split}
&\left(\frac{\partial^2}{\partial r^2} +
\frac{2}{r}\frac{\partial}{\partial r} + \frac{1}{r^2} \frac{\partial^2}{\partial \theta^2}
+ \frac{\cos \theta}{r^2 \sin\theta}\frac{\partial}{\partial \theta}
+ 2 \mu^2 (1-X^2) - m^2 Y^2
 \right) X = 0\\
&\left(\frac{\partial^2}{\partial r^2} +
\frac{2}{r}\frac{\partial}{\partial r} + \frac{1}{r^2} \frac{\partial^2}{\partial \theta^2}
+ \frac{\cos \theta}{r^2 \sin\theta}\frac{\partial}{\partial \theta}
-  m^2 X^2 + \omega^2 -\frac{n^2}{r^2 \sin^2\theta}
 \right) Y = 0
\end{split}
\label{eqs-spin}
\ee
Note that these equations are symmetric with respect to reflections in the $\theta=\pi/2$ plane.

To find numerical solutions of these coupled partial differential equations
we used the software package CADSOL based on the Newton-Raphson algorithm \cite{schoen}.
The numerical calculations are mainly performed on an equidistant grid
in spherical coordinates $r$ and $\theta$. Typical grids we used have sizes $70 \times 60$.
As before, we map the infinite interval of the variable $r$ onto the compact radial coordinate
$x=\frac{r/r_0}{1+r/r_0}  \in [0:1]$. Here $r_0$ is a real scaling constant, which typically is taken as $r_0 = 4 - 6$.
For spinning Q-balls the component $Y(r,\theta)$ must vanish at the origin,
both in the massive and in the massless cases. The restriction of regularity also yields $\partial_r X(r\,\theta)=0$
as $r\to 0$.

To secure the condition of regularity on the symmetry axis we
impose there the boundary conditions
\be
\partial_\theta X\bigl.\bigr|_{\theta=0,\pi}=0  \, , \qquad Y\bigl.\bigr|_{\theta=0,\pi}=0 \, .
\ee

\begin{figure}[h!]
\begin{center}
\setlength{\unitlength}{0.1cm}
\includegraphics[height=.26\textheight, angle =0]{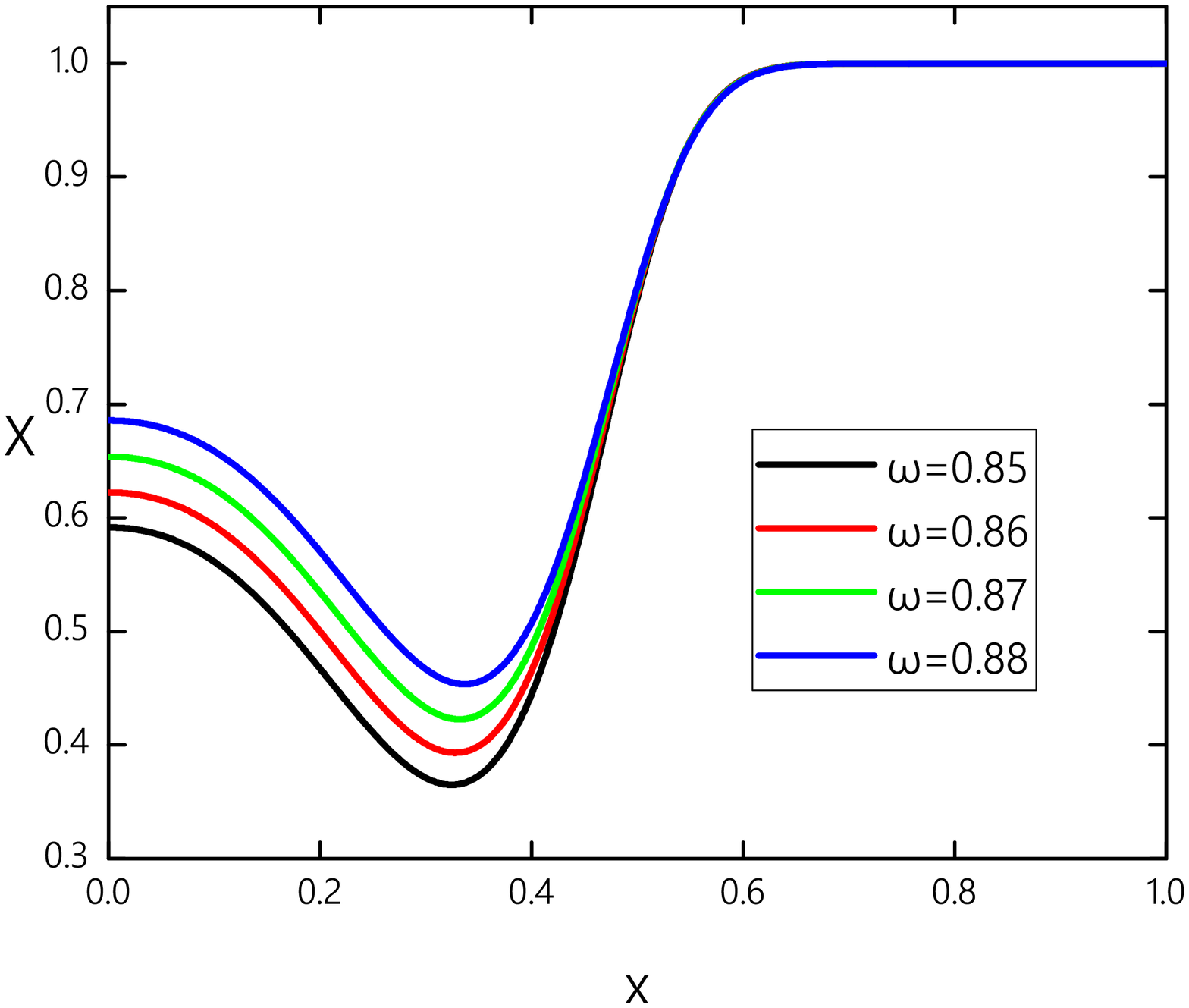}
\includegraphics[height=.26\textheight, angle =0]{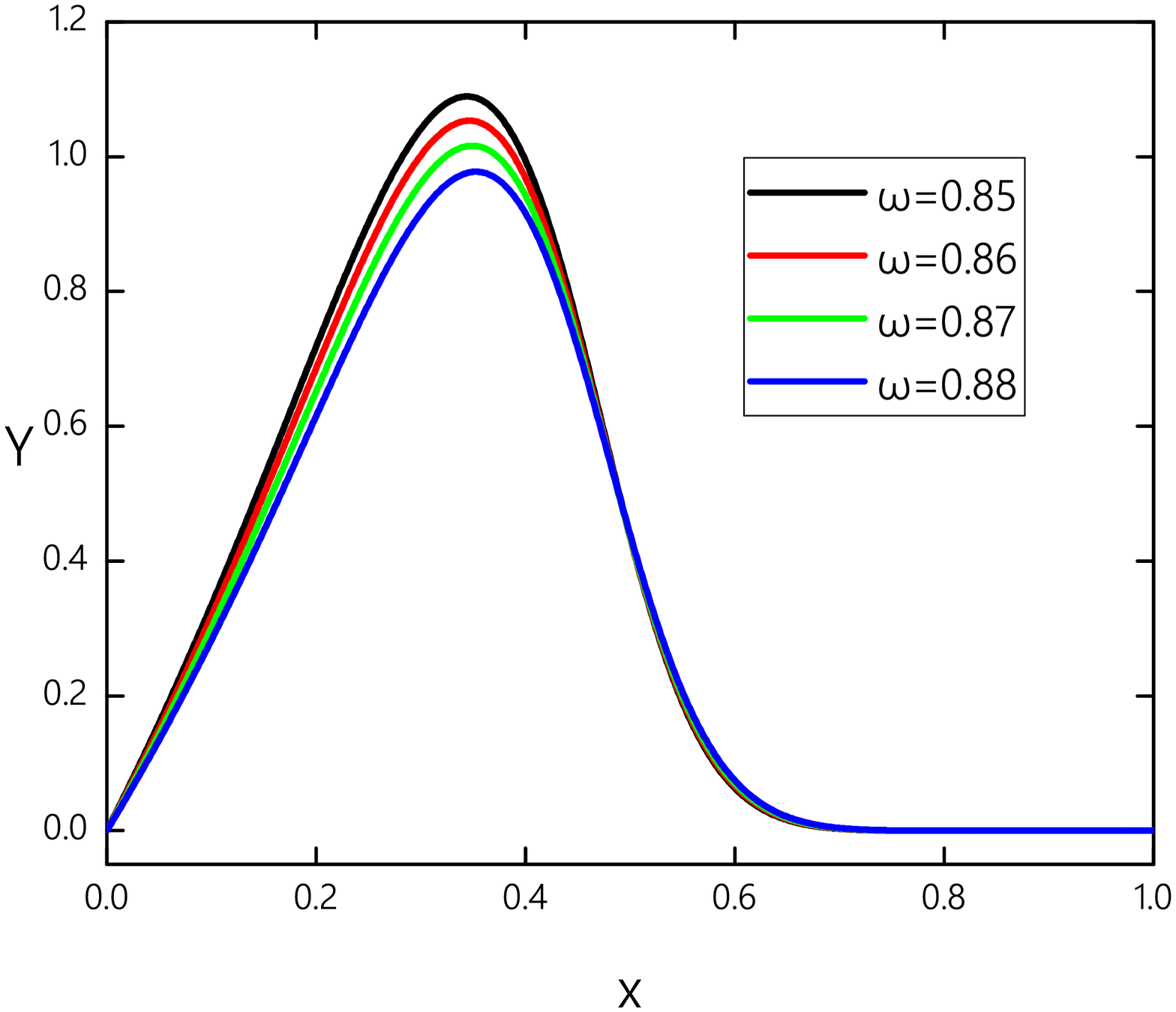}
\includegraphics[height=.26\textheight, angle =0]{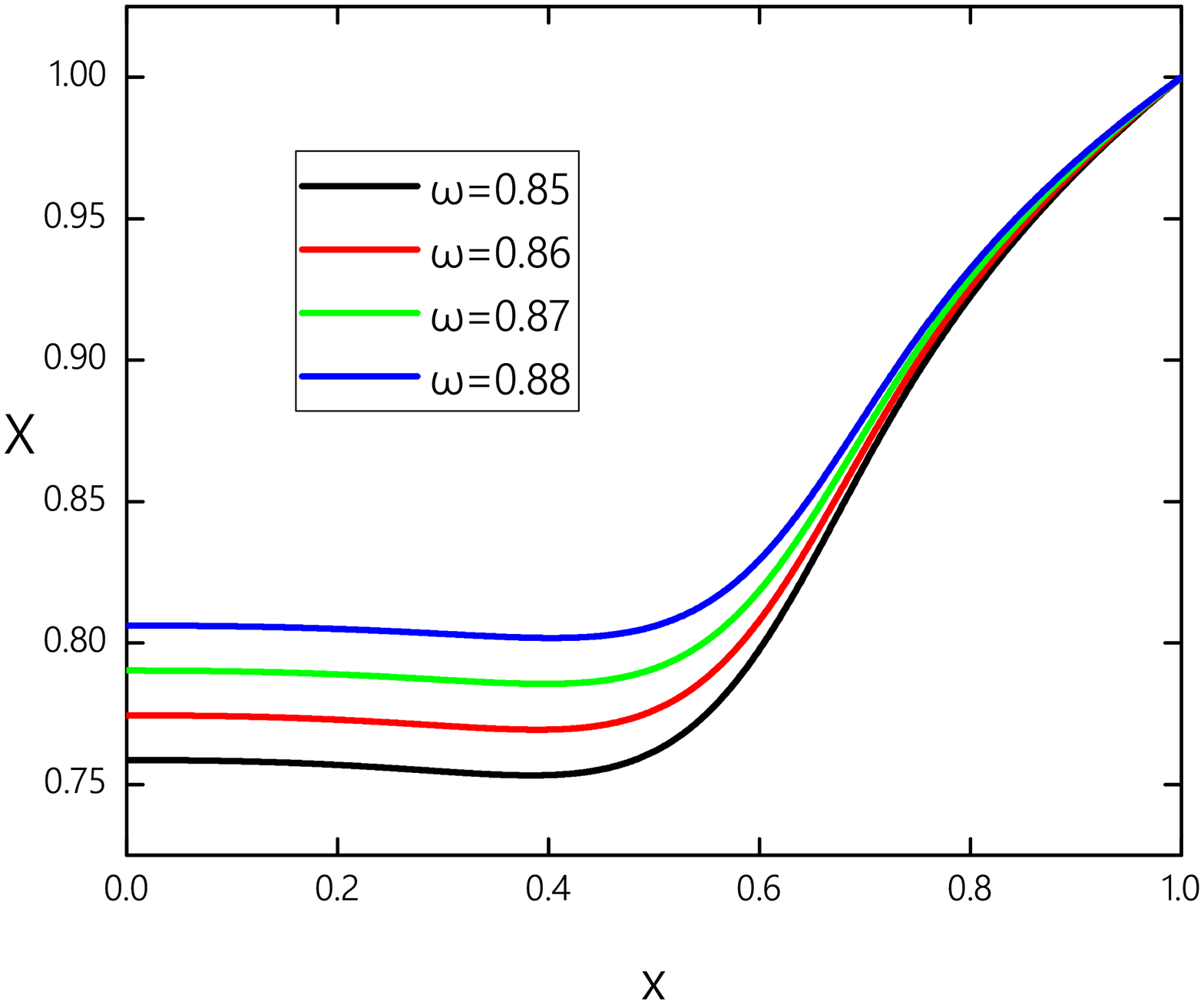}
\includegraphics[height=.26\textheight, angle =0]{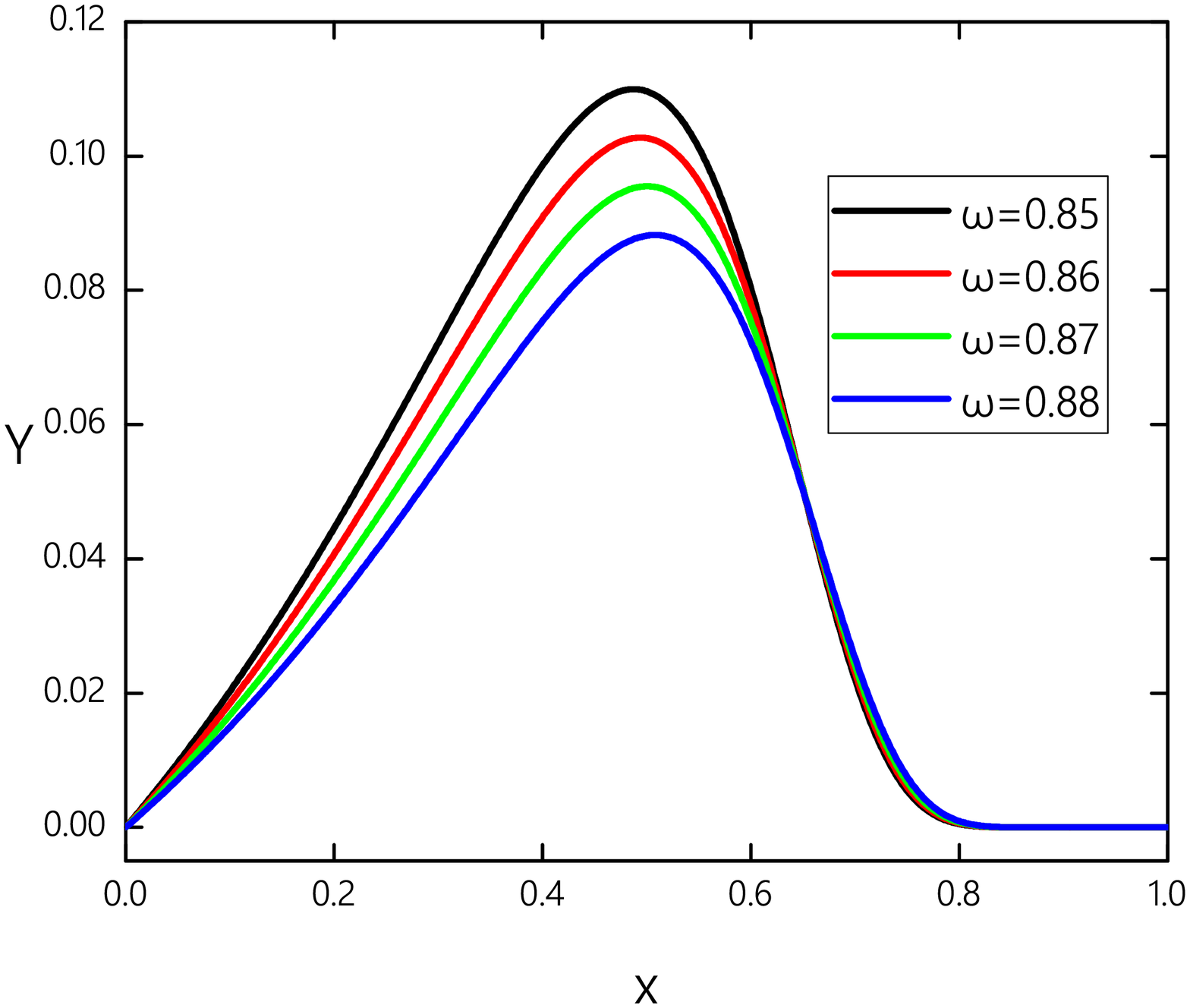}
\end{center}
\caption{The profiles of the field components of the spinning $n=1$ Q-balls with positive parity
are plotted as functions of the compact radial variable $x=\frac{r/r_0}{1+r/r_0}$ at $r_0=6$, $\theta=\pi/2$ for some set of values
of angular frequency $\omega$ at $m=1$ and $\mu^2=0.25$ (upper plots) and $\mu^2=0$ (bottom plots).}
\label{fig4}
\end{figure}

The spinning Q-balls correspond to stationary points of the action functional,
they exist only for a restricted frequency range. Previously they have been
constructed only for the model with a single complex field and sextic potential \cite{Volkov:2002aj,Kleihaus:2005me,Kleihaus:2007vk}.
A peculiar feature of these Q-balls is that
for a non-zero rotational quantum number $n$ there are
two different solutions with even and odd parity. The corresponding
energy density distribution forms one or more tori, respectively. Similar to the spherically symmetric configuration with $n=0$,
the spinning Q-balls exist for all range of values of the angular frequency $\omega \in [0,1]$.
The limit of small Q-balls corresponds to the  $\omega$ approaching the upper critical value, given by the mass of the complex component, in the
opposite limit of small value of angular frequencies, spinning Q-ball rapidly expands.

We observe that, similar to the spinning Q-balls in the single component model \cite{Volkov:2002aj,Kleihaus:2005me,Kleihaus:2007vk}
for each value of integer winding number $n$, there are two types of solutions possessing
different parity, so called parity-even and parity-odd Q-balls.
Indeed, the equations \re{eqs-spin} in the limiting case of $\omega^2 \sim m^2$ (small Q-balls)
can be linearized. Then the second of these equations is reduced to the standard harmonic equation and the solution are associated
with the usual spherical harmonics
$$
Y_l^n(\theta, \varphi)=\sqrt{\frac{2l+1}{4\pi} \frac{(l-n)!}{(l+n)!}}P_l^n(\cos\theta)e^{in\varphi} \, .
$$
Here $P_l^n(\cos\theta)$ are
the associated Legendre polynomials of degree $l$ and order $n$. Thus, the spherically symmetric fundamental  Friedberg-Lee-Sirlin
Q-ball corresponds
to the spherical harmonic $Y_0^0$, and there are two spinning configuration in the sector $n=1$,
the parity-even solution $Y_1^1$ and parity-odd solution $Y_2^1$, respectively. Further, this observation suggests that the
equations \re{eqs-spin} also support solutions which correspond to higher energy angular excitations
of the fundamental Q-ball \cite{Brihaye:2007tn}.

\begin{figure}[h!]
\begin{center}
\setlength{\unitlength}{0.1cm}
\includegraphics[height=.34\textheight, angle =-90]{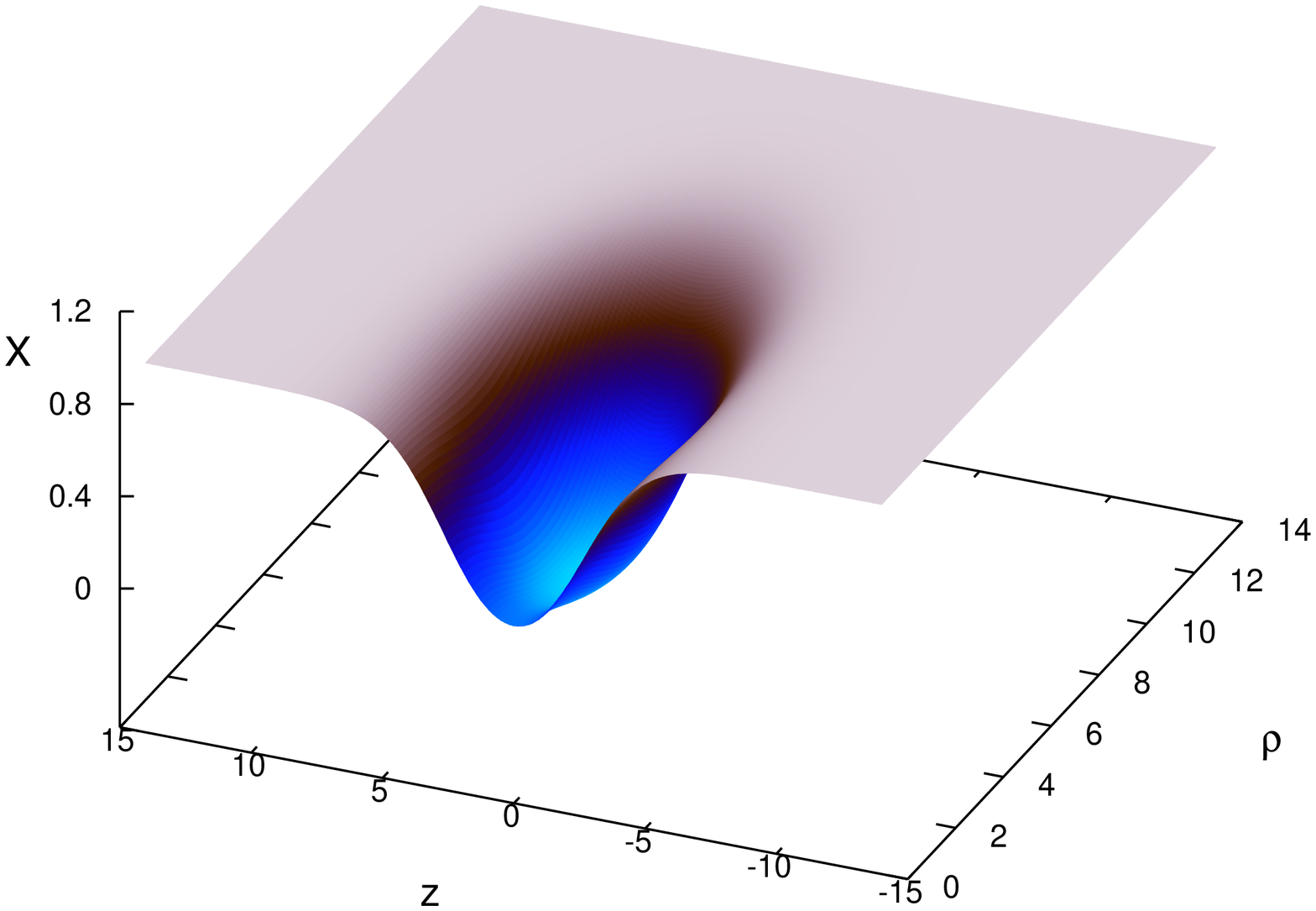}
\includegraphics[height=.34\textheight, angle =-90]{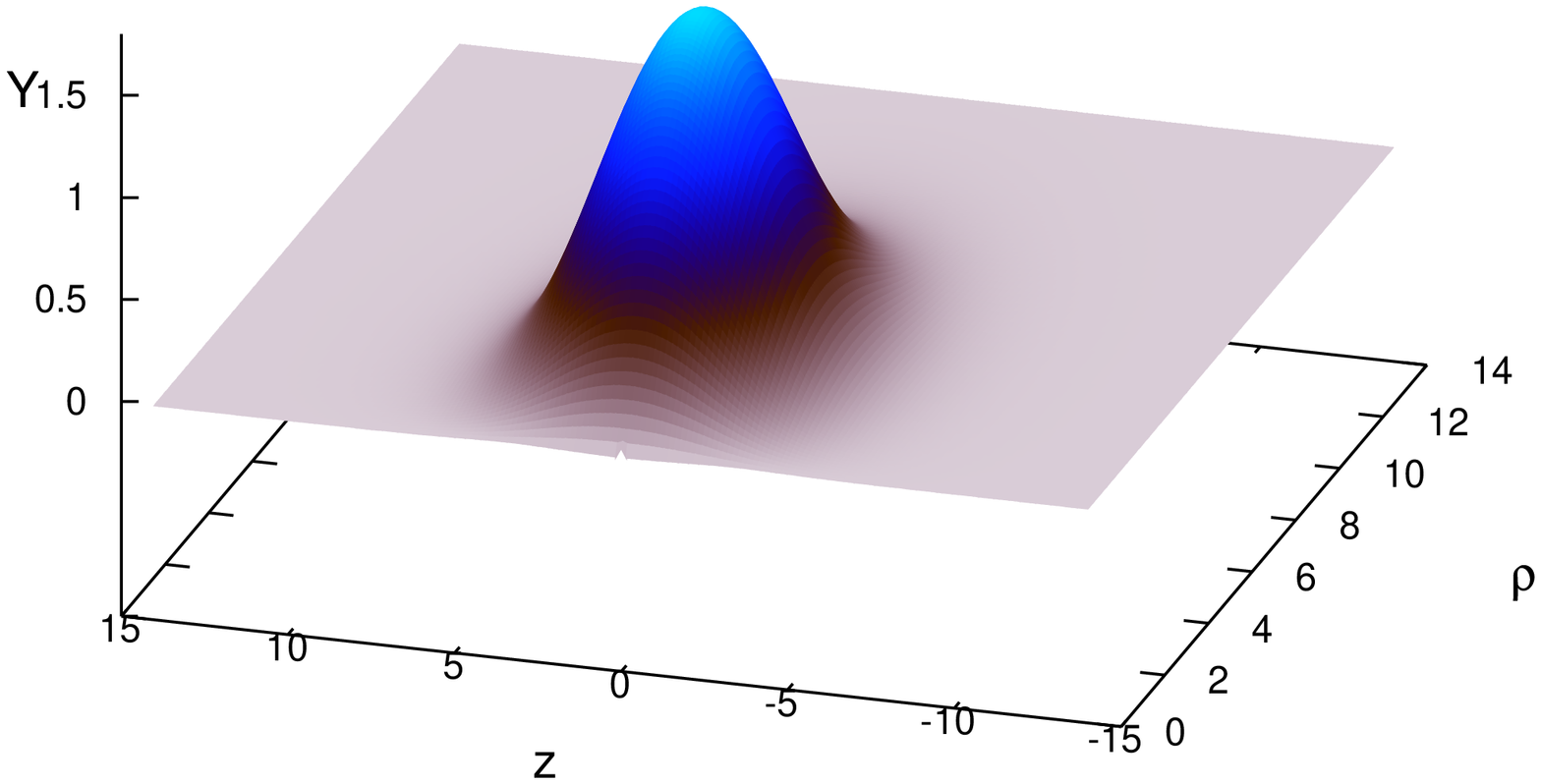}
\includegraphics[height=.34\textheight, angle =-90]{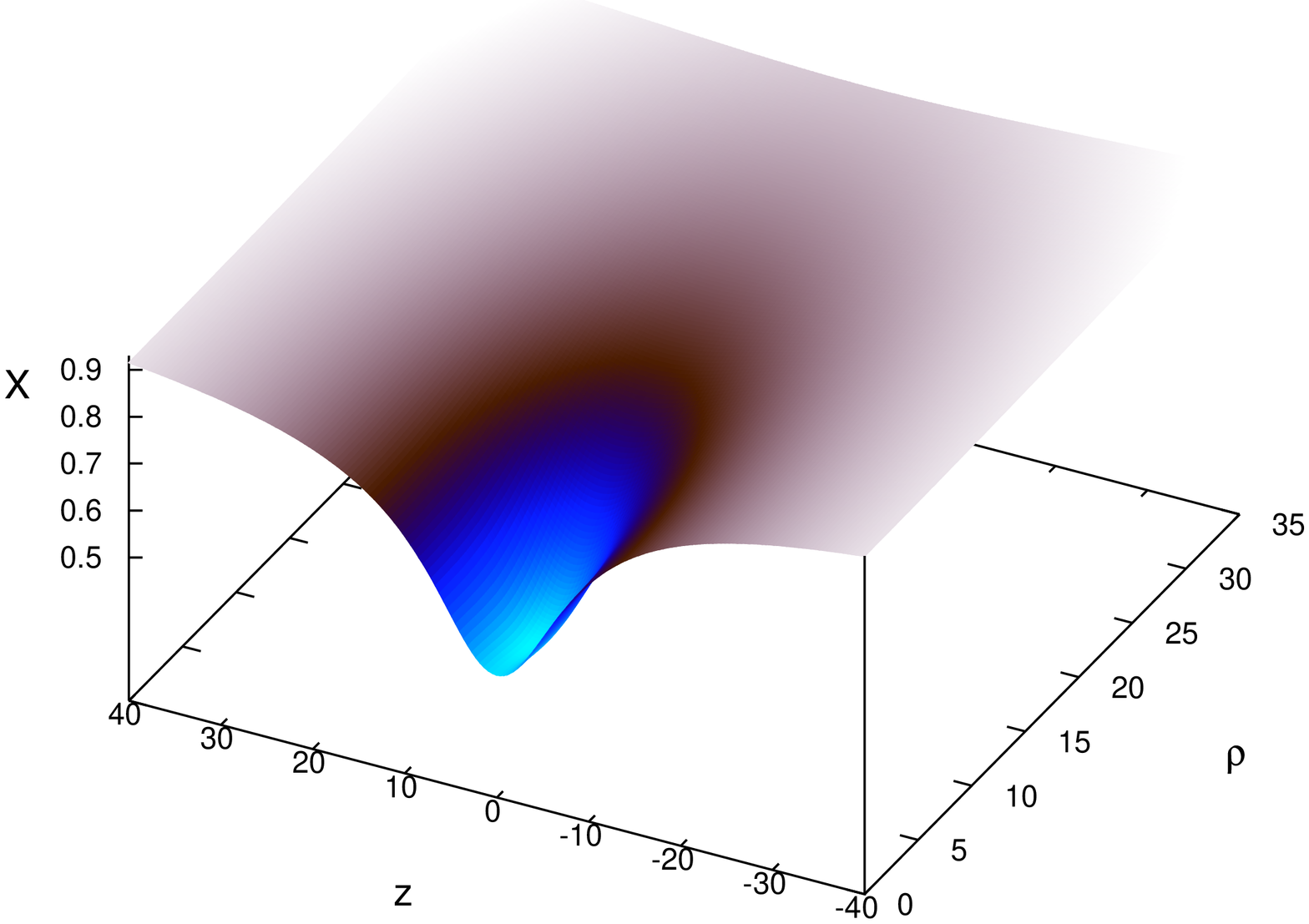}
\includegraphics[height=.34\textheight, angle =-90]{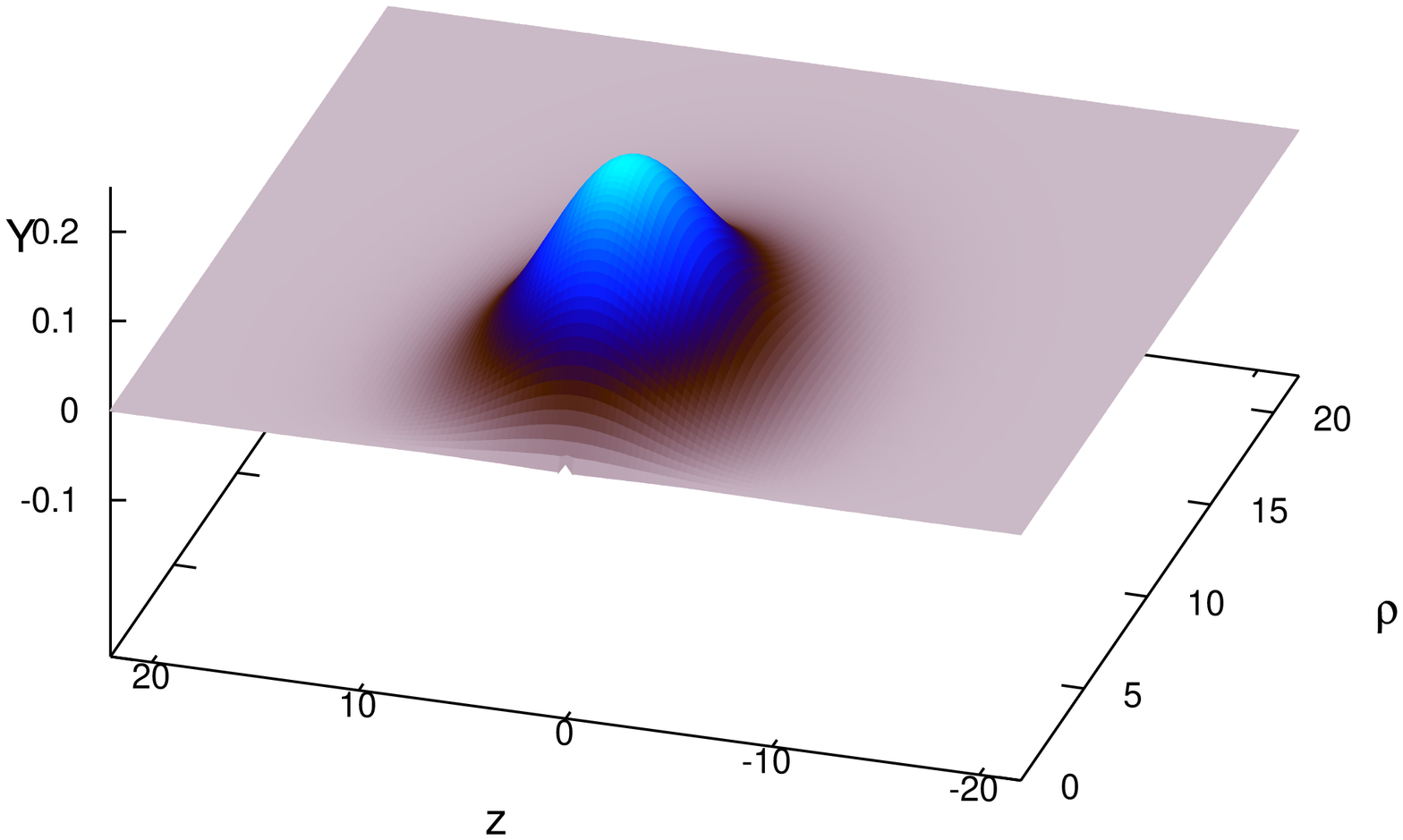}
\end{center}
\caption{The field components of the spinning $n=1$ parity-even Q-ball at
$\omega=0.65$, $m=1$, $\mu^2=0.25$ (upper row) and $\mu^2=0$ (bottom row) are shown as functions of the coordinates $\rho = \sin \theta$ and
$z=r\cos \theta$.
}
\label{fig5}
\end{figure}

\begin{figure}[h!]
\begin{center}
\setlength{\unitlength}{0.1cm}
\includegraphics[height=.34\textheight, angle =-90]{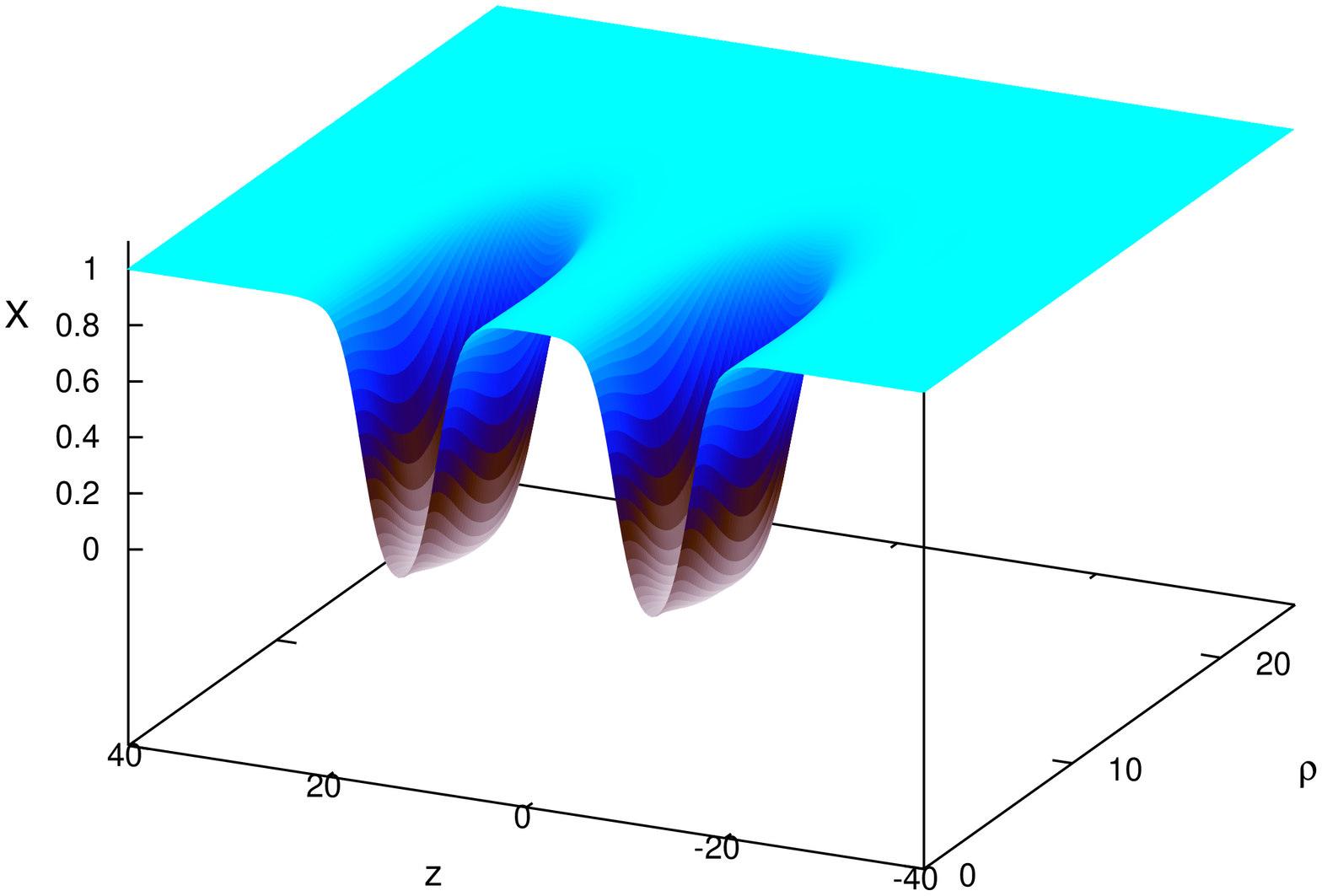}
\includegraphics[height=.34\textheight, angle =-90]{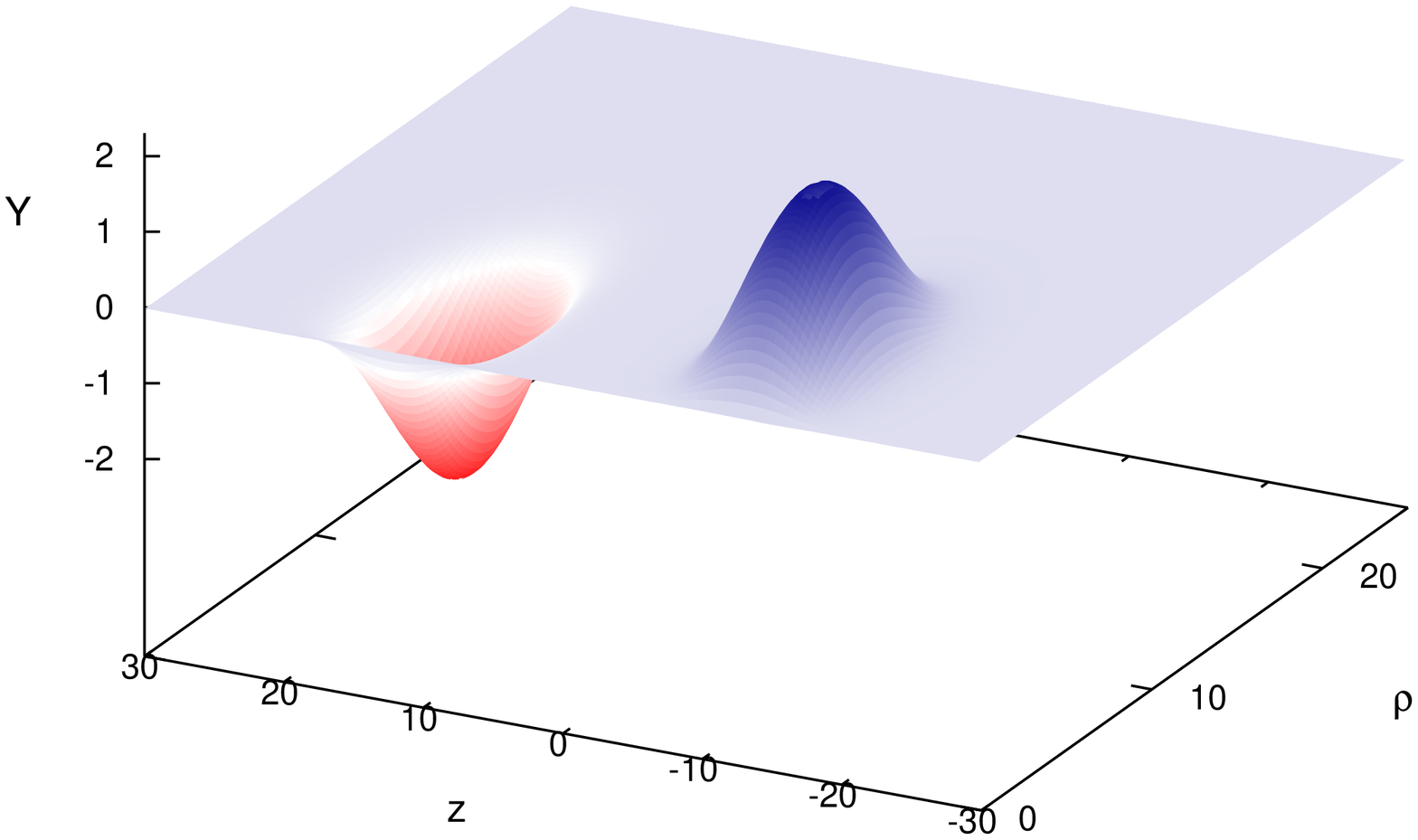}
\includegraphics[height=.34\textheight, angle =-90]{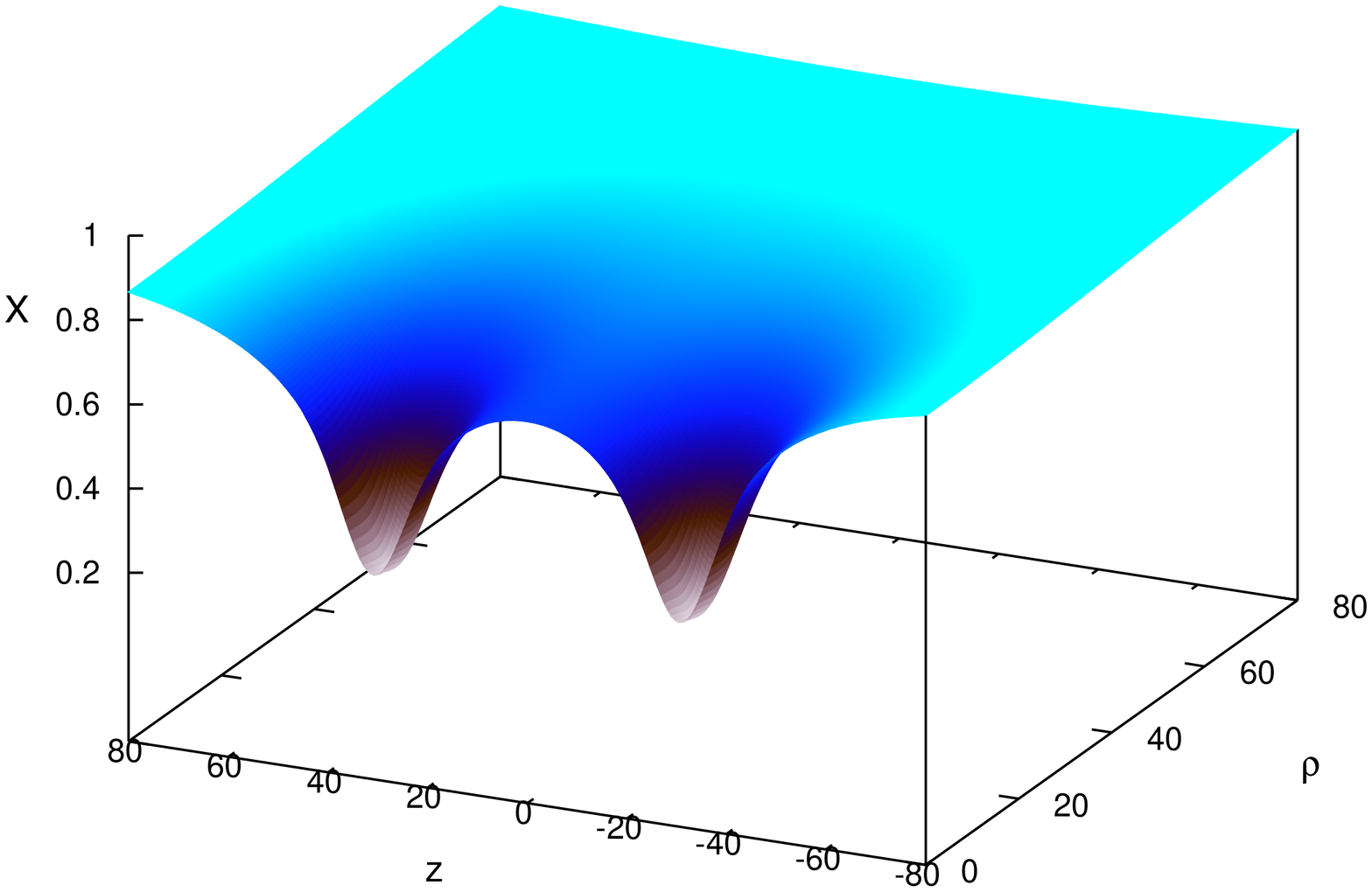}
\includegraphics[height=.34\textheight, angle =-90]{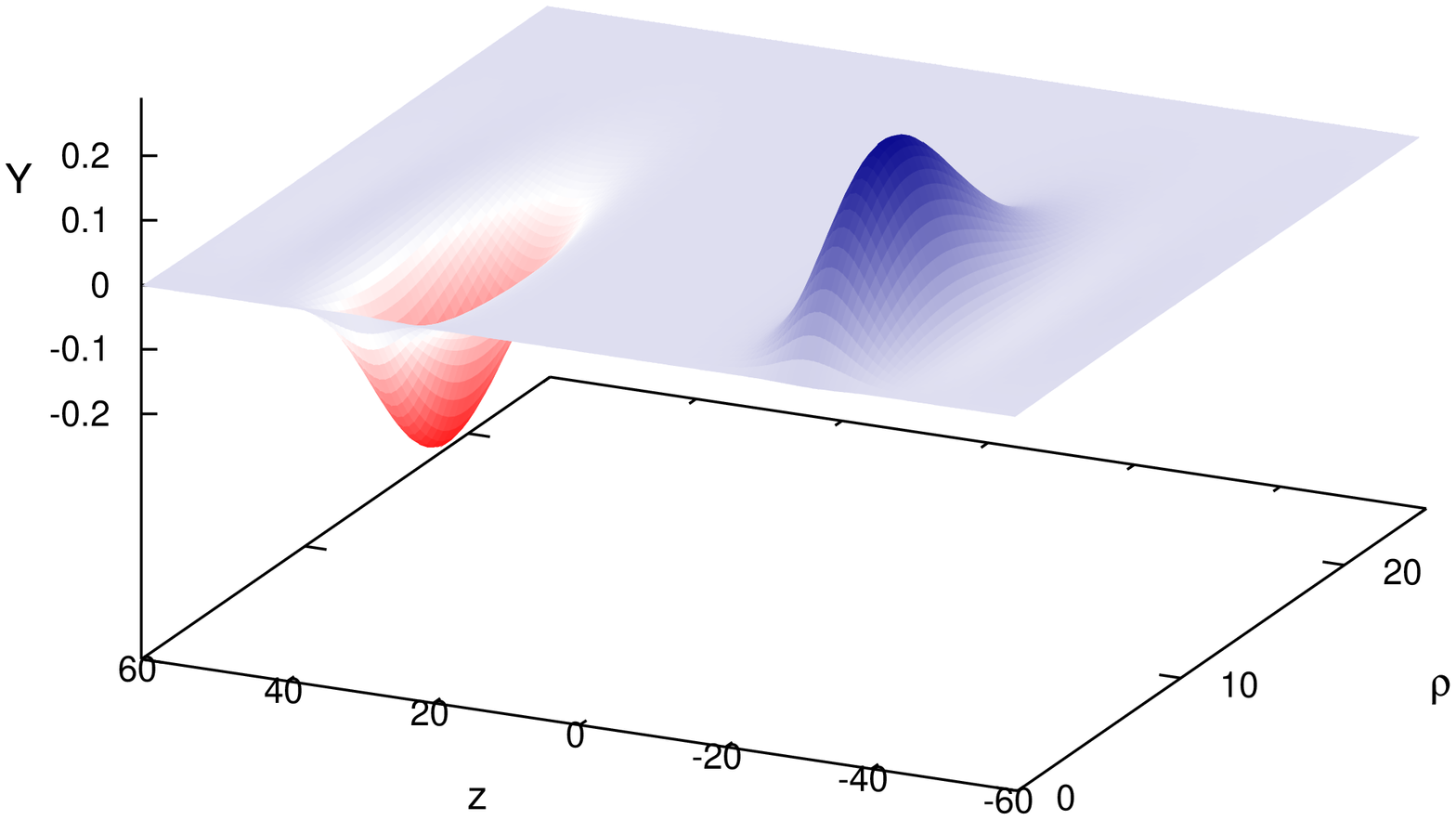}

\end{center}
\caption{The field components of the spinning $n=1$ parity-odd Q-ball at
$\omega=0.50$, $m=1$, $\mu^2=0.25$ (upper row) and $\mu^2=0$ (bottom row) are shown as functions of the coordinates
$\rho = \sin \theta$ and
$z=r\cos \theta$.
}
\label{fig6}
\end{figure}

In Figs.~\ref{fig5},\ref{fig6} we displayed the fields of the spinning parity-even and parity-odd Q-balls, both in the
massive and massless cases. For parity-even  solutions the spinning component $Y(r,\theta)$  is maximal in the symmetry plane, as $\mu=0$
the massless component $X(r,\theta)$ is minimal at the origin, it decays asymptotically, according to \re{massless-asymp}
(see Fig.~\ref{fig5}, bottom left plot). If the mass of the $X(r,\theta)$ component is non-zero,  it decays exponentially, as
it is seen in Fig~\ref{fig5}, upper left plot. We also observe that in the latter case the minimum of this component is shifted to the
$x-y$ plane. The energy density distributions of the rotating even-parity Q-balls is torus-shaped in both cases.
However, as the component $X(r,\theta)$ is massive, for the same value of the frequency $\omega$, both the energy and
the amplitudes of the fields are much larger.

Fig.~\ref{fig4} exhibits the profiles of the corresponding positive parity solutions in the symmetry plane
$\theta=\pi/2$ at $m=1$, $\mu^2=0.25$ and for $\mu=0$ for some set of values of the angular frequency $\omega$.

\begin{figure}[h!]
\begin{center}
\setlength{\unitlength}{0.1cm}
\includegraphics[height=.26\textheight, angle =0]{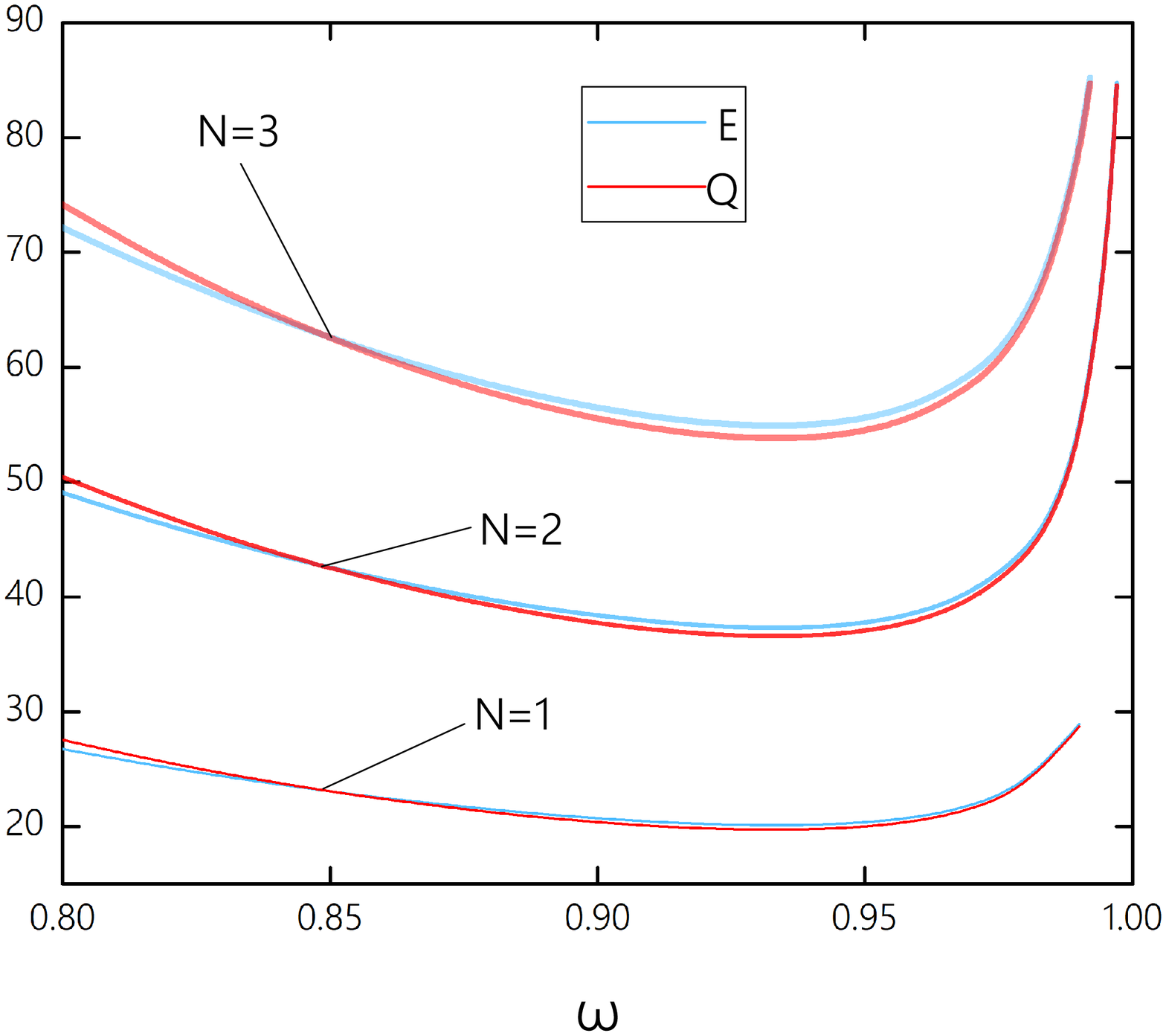}
\includegraphics[height=.26\textheight, angle =0]{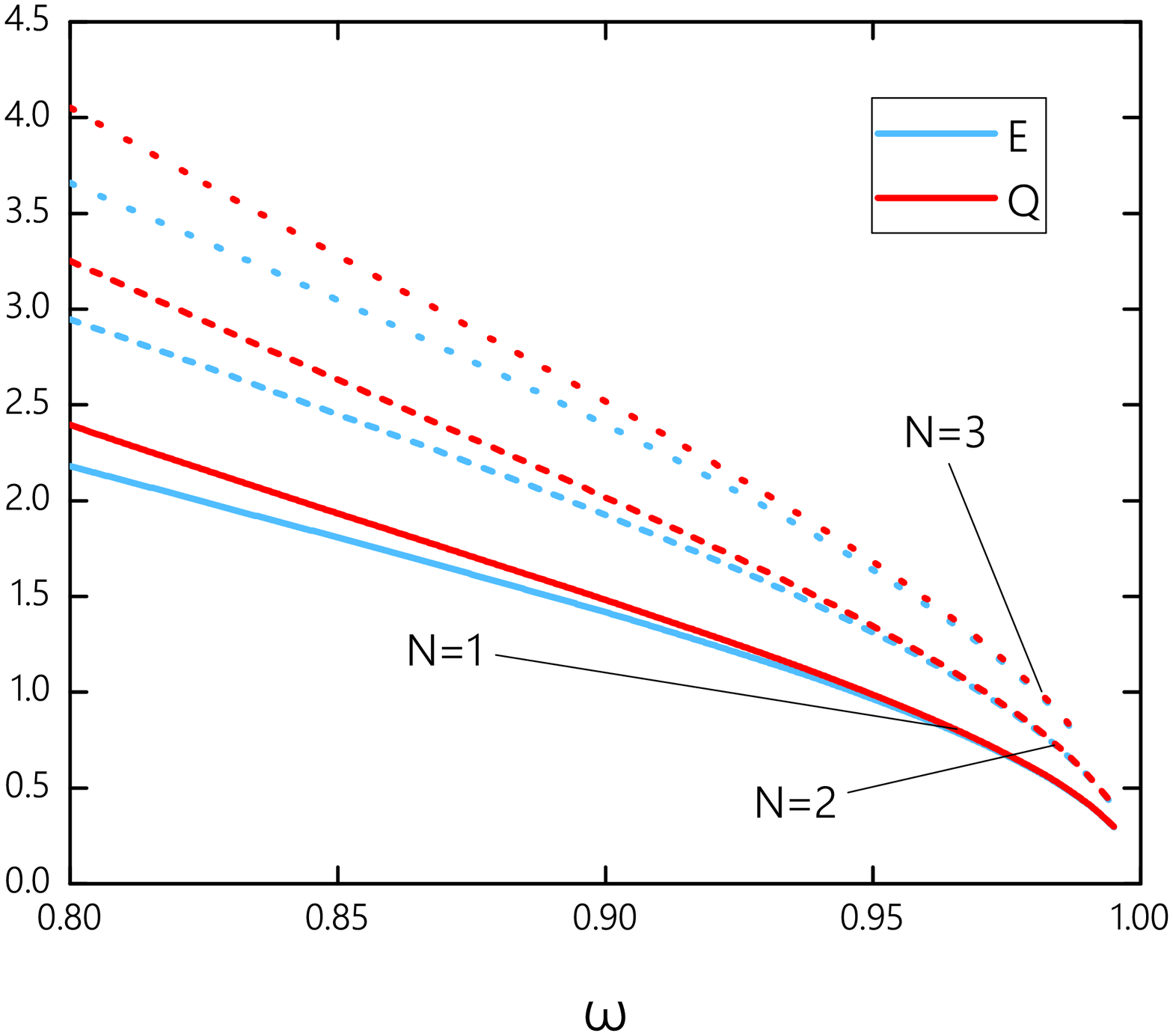}
\end{center}
\caption{The energy and the charge of the spinning parity-even Q-balls are shown in units of $8\pi$ as functions
of the angular frequency $\omega$ at $m=1$, $\mu^2=0.25$ (left) and $\mu^2=0$ (right).
}
\label{fig7}
\end{figure}

The field components of the parity-odd $n=1$ Q-ball are presented in Fig.~\ref{fig6}.
For this solutions the spinning component $Y(r,\theta)$ vanishes in the equatorial plane while the real field
$X(r,\theta)$ possess two minima, located symmetrically with respect to the $x-y$ plane. The difference
between the massive and massless cases is that, as $\mu =0$, the field $X(r,\theta)$ decays asymptotically as $\sim 1/r$.
Further, in such a case the minima of this component are located on the $z$-axis, see Fig.~\ref{fig6}, bottom left plot.
The energy density distribution of the parity-odd Q-balls has a double torus structure. Note that as the real component $X(r,\theta)$
becomes massless, the characteristic size of the configuration grows, the tori are well separated from each other.

Considering the frequency dependence of rotating Q-balls, we found that it is qualitatively the same as in
the case of the fundamental $n=0$ solutions, see Figs.~\ref{fig2},\ref{fig3} above.

In Fig.~\ref{fig7}, left plot, we show the charge and the energy of the parity-even Q-balls as functions of the angular
frequency $\omega$. We observe that the solutions exist for all allowed range of values of $\omega$ restricted from above by the
mass of the complex field $m=1$. As $\omega$ approaches this upper limit, the amplitude of both components is decreasing.
As the real component $X(r,\theta)$ remains massive,
it decays exponentially. In such a case
we observe the same usual pattern as in the model with single component
complex field and a polynomial self-interaction potential \cite{Volkov:2002aj,Kleihaus:2005me,Kleihaus:2007vk}. Both the energy and the charge of the
configuration of these Q-balls are minimal at some critical value $\omega_{cr}$ and the curve $E(Q)$ shows the same cusp structure as
displayed in Fig.~\ref{fig3}, left plot. Thus, there are two branches of $E(Q)$ curves, the existence of
two different solutions with the same value of charge $Q$ indicates that the more energetic
configurations on the upper branch are unstable.

The situation is different for the spinning Q-balls with massless long-range component $X(r,\theta)$. As for the spherically symmetric
$n=0$ configurations,  both the charge and the energy monotonically depend on $\omega$,
see Fig.~\ref{fig3}, right plot. Thus, there is no critical frequency and only one branch of classically stable solutions exist for all range
of values of $\omega$. The scalar charge $C$ of the spinning Q-balls with massless hair also depends on the angular frequency.
Numerical calculations show that dependency of its value per rotational quantum number $n$ is identical with the dependency of the scalar charge
of the spherically symmetric hairy Q-ball displayed in Fig.~\ref{fig8} above.

\section{Conclusions}
The objective of this work was to investigate properties of new Q-ball solutions with long-range massless scalar hair,
existence of which was conjectured in the pioneering study \cite{Levin:2010gp}.

As the real scalar component of the Friedberg-Lee-Sirlin Q-ball remains massive, these solutions
exhibit the same general pattern as the corresponding non-topological solitons in the non-renormalizable model with polynomial
potential. The solutions exist only in a finite frequency range, which is restricted from above by the value of the
mass of the complex component. We also found that there are spinning generalizations of the fundamental spherically symmetric
Friedberg-Lee-Sirlin Q-ball with both even and odd parity. The large Q-balls correspond the the small values of the
angular frequency, they tend to the corresponding states of the perturbative spectrum  as $\omega$ approaches the upper limiting value.

However, this pattern drastically changes, as the potential is vanishing.
The vacuum expectation value of the real massless scalar field still remains non zero,
this component possess long range Coulomb asymptotic tail.
Such a hairy Q-ball has an additional scalar charge $C$, which corresponds to the weak long-range attractive interaction between the solitons.
On the other hand, there is a short-range Yukawa interaction mediated by the complex component of the  Friedberg-Lee-Sirlin Q-ball.
If the interacting Q-balls are in phase, this interaction is attractive,
while if they are  out-of-phase it is  repulsive, thus the resulting pattern of interaction between the Q-balls is rather complicated.

We found that the branch structure of the hairy Q-balls is different from the case when the real component remains massive.
Notably, both the energy and the charge of the configuration decreases monotonically as $\omega$ increases.
There is no second unstable branch of solutions, the hairy Q-balls with long-range real scalar component
are classically stable within all range of values of the angular frequency.

The work here should be taken further by considering gauged spinning Q-balls with massless real scalar component, another
interesting direction is to investigate properties of self-gravitating spinning Q-balls without potential.
We hope we can address these issues in our future work.

\section*{Acknowledgements}
Y.S. gratefully acknowledges support from the Ministry of Education and Science
of Russian Federation, project No 3.1386.2017. He would like to thank Nobuyuki Sawado and the
Department of Physics, Tokyo University of Science, for its kind hospitality.


\begin{thebibliography}{00}
\bibitem{Friedberg:1976me} R.~Friedberg, T.D.~Lee and A.~Sirlin,
Phys.\ Rev.\ D {\bf 13} (1976)  2739.
\bibitem{Coleman:1985ki}S.R.~Coleman,
Nucl.\ Phys.\ B  {\bf 262}  (1985)  263; Erratum: Nucl.\ Phys.\ B  {\bf 269} (1986)   744
\bibitem{Friedberg:1986tq} R.~Friedberg, T.~D.~Lee and Y.~Pang,
Phys.\ Rev.\ D  {\bf 35} (1987)  3658.
\bibitem{Jetzer:1991jr} P.~Jetzer,
Phys.\ Rept.\  {\bf 220}  (1992) 163.
\bibitem{Lee:1986ts}T.~D.~Lee,
 Phys.\ Rev.\ D  {\bf 35}  (1987) 3637.
\bibitem{Kuzenko}A.~Kusenko,
Phys.\  Lett.\ B  {\bf 405}   (1997)  108.
\bibitem{Affleck}I.~Affleck and M.~Dine,
Nucl.\ Phys.\ B  {\bf 249}  (1985)  361.
\bibitem{Kusenko:1997si}A.~Kusenko and M.~E.~Shaposhnikov,
Phys.\ Lett.\ B  {\bf 418}  (1998)  46.
\bibitem{Anagnostopoulos:2001dh} K.~N.~Anagnostopoulos, M.~Axenides, E.~G.~Floratos and N.~Tetradis,
Phys.\ Rev.\ D  {\bf 64} (2001)  125006.
\bibitem{Lee:1988ag}K.M.~Lee, J.A.~Stein-Schabes, R.~Watkins and L.M.~Widrow,
Phys.\ Rev.\ D  {\bf 39}  (1989) 1665.
\bibitem{Kusenko:1997vi}A.~Kusenko, M.E.~Shaposhnikov and P.G.~Tinyakov,
Pisma Zh.\ Eksp.\ Teor.\ Fiz.\  {\bf 67} (1998)  229;
  [JETP Lett.\  {\bf 67}, 247 (1998)]
\bibitem{Safian:1987pr} A.~M.~Safian, S.R.~Coleman and M.~Axenides,
Nucl.\ Phys.\ B  {\bf 297} (1988)   498.
\bibitem{Enqvist}K.~Enqvist and M.~Laine,
JCAP 0308  (2003)  003.
\bibitem{Bunkov}Yu.M.~Bunkov and G.E.~Volovik,
Phys.\ Rev.\ Lett.\  {\bf 98}  (2007)  265302.
\bibitem{Levin:2010gp}
  A.~Levin and V.~Rubakov,
  Mod.\ Phys.\ Lett.\ A {\bf 26} (2011) 409.
\bibitem{Volkov:2002aj}M.S.~Volkov and E.~Wohnert,
Phys.\ Rev.\  D  {\bf 66} (2002)  085003.
\bibitem{Kleihaus:2005me}B.~Kleihaus, J.~Kunz and M.~List,
Phys.\ Rev.\ D  {\bf 72} (2005)  064002 .
\bibitem{Kleihaus:2007vk}
  B.~Kleihaus, J.~Kunz, M.~List and I.~Schaffer,
  Phys.\ Rev.\ D {\bf 77} (2008) 064025
\bibitem{Radu:2008pp}E.~Radu and M.S.~Volkov,
Phys.\ Rept.\  {\bf 468}  (2008)  101.
\bibitem{Brihaye:2007tn}Y.~Brihaye and B.~Hartmann,
Nonlinearity  {\bf 21}  (2008) 1937.
\bibitem{shnir}Y.M.~Shnir, \textit{Topological and Non-Topological Solitons in Scalar Field Theories},
(Cambridge University Press, Cambridge, 2018)
\bibitem{schoen}
W.~Sch\"onauer  and R.~Wei\ss,
"Efficient vectorizable PDE solvers"
J. Comput. Appl. Math. 1989. V. 27. P. 279\\
M.~Schauder, R.~Wei\ss and  W.~Sch\"onauer,
"The CADSOL Program Package",
Universit\"at Karlsruhe, 1992. Interner Bericht Nr. 46/92.



\end{thebibliography}
\end{document}